\documentclass[article,twocolumn,showpacs,superscriptaddress]{revtex4-1}

\usepackage{graphicx}
\graphicspath{{./},{figures/}}

\usepackage{amssymb}
\usepackage{bm}
\usepackage{braket}
\usepackage{amsmath}
\usepackage{tabularx}
\usepackage{ltablex}
\usepackage{multirow}
\usepackage[dvipsnames]{xcolor}

\usepackage[colorlinks=true,
                linkcolor=blue,
	        citecolor=blue,
	        urlcolor=blue]{hyperref}

\newcommand\2{$_2$}
\newcommand\3{$_3$}

\newcommand{\dqmp}{Department of Quantum Matter Physics, University of Geneva,CH-1211 Geneva, Switzerland}
\newcommand{\dtp}{Department of Theoretical Physics, University of Geneva, CH-1211 Geneva, Switzerland}
\newcommand{\unimore}{Dipartimento di Scienze Fisiche, Informatiche e Matematiche, University of Modena and Reggio Emilia, IT-41125 Modena, Italy}

\AtBeginDocument{\usepackage{booktabs}}
\begin{document}

\title{Gate-tunable imbalanced Kane-Mele model in encapsulated bilayer jacutingaite}

\date{\today}
\author{Louk Rademaker}
\affiliation{\dtp}
\author{Marco Gibertini}
\affiliation{\unimore}
\affiliation{\dqmp}

\begin{abstract}
We study free, capped and encapsulated bilayer jacutingaite Pt\2HgSe\3 from first principles. While the free standing bilayer is a large gap trivial insulator, we find that the encapsulated structure has a small trivial gap due to the competition between sublattice symmetry breaking and sublattice-dependent next-nearest-neighbor hopping. Upon the application of a small perpendicular electric field, the encapsulated bilayer undergoes a topological transition towards a quantum spin Hall insulator. We find that this topological transition can be qualitatively understood by modeling the two layers as uncoupled and described by an imbalanced Kane-Mele model that takes into account the sublattice imbalance and the corresponding inversion-symmetry breaking in each layer. Within this picture, bilayer jacutingaite undergoes a transition from a 0+0 state, where each layer is trivial, to a 0+1 state, where an unusual topological state relying on Rashba-like spin orbit coupling emerges in only one of the layers.

\end{abstract}

\maketitle

\section{Introduction}

Topological insulators have a finite gap in their bulk energy spectrum, but differ from standard (trivial) insulators because a non-zero topological invariant is associated with the manifold of occupied states~\cite{review_hasankane_2010,bernevig_topological_2013}. In most cases, the non-trivial topological invariant results in the appearance of metallic states that cross the bulk gap close to the boundary of a finite-size system. The nature of the topological invariant depends on the dimensionality and the underlying fundamental symmetries of the system~\cite{Kitaev2009,Ryu2010}, including crystal symmetries~\cite{Fu2011,Slager2013}. A paradigmatic example is the integer quantum Hall state in two dimensions (2D), for which the topological invariant is an integer $C$ --known as Chern number-- which provides the number of (chiral) states localized close to each edge and is related to the quantized Hall conductivity $\sigma_{xy}= C\ e^2/h$~\cite{TKNN,Kohmoto1985}.

When time-reversal symmetry is preserved  in 2D, although the Chern number vanishes identically, another topological invariant $\nu$ can be introduced~\cite{kane_quantum_2005,kane_z2_05,Bernevig2006,bernevig_quantum_2006}, which is a $\mathbb{Z}_2$ number that can assume only two values: 0 or 1, i.e.\ trivial or non-trivial. 
As a consequence of time-reversal symmetry, gapless states appear at the edges of the system in pairs of counter-propagating (helical) modes and a bulk-boundary correspondence relates $\nu$ to the parity of the number of such pairs. In particular, we  have that an even number of helical pairs  is topologically trivial ($\nu=0$), as states belonging to different pairs can be mixed and adiabatically gapped out without breaking time-reversal symmetry. On the contrary, time-reversal-invariant topological insulators ($\nu=1$), also known as quantum spin Hall insulators (QSHIs), have an odd number of pairs, so that the presence at each edge of at least one pair of helical gapless states is robust. 

Experimental realizations of the QSHI phase have been reported in semiconductor quantum wells based on HgTe/CdTe~\cite{Konig2007,Roth2009,Grabecki2013,Konig2013} and InAs/GaSb~\cite{ Knez2011,Suzuki2013} heterostructures, as well as in 2D materials like WTe\2~\cite{Fei2017,Tang2017,Wu2018,Shi2019}. In all these systems, the operating conditions where transport is dominated by edge states are limited to fairly low temperatures owing to their small bulk energy gap. A breakthrough could be represented by monolayer Pt\2HgSe\3, which has been predicted using first-principles simulations to be the first materials realization of the seminal Kane-Mele model~\cite{kane_quantum_2005,kane_z2_05} for QSHIs, with a substantial energy gap of 0.5~eV~\cite{Marrazzo2018} (and could even give rise to a Chern insulator when functionalized~\cite{Luo2021} or interfaced with a magnetic material like CrI\3~\cite{Liu2020}). Although monolayers of this material could be potentially exfoliated~\cite{Mounet2018} from a bulk layered mineral called jacutingaite~\cite{cabral_first_obs_08,jacutingaite_exp_12}, a clear experimental validation is still lacking~\cite{Kandrai2020}.

When two QSHI monolayers are stacked together to form a bilayer, the system is expected to become trivial in the limit of weak interlayer coupling. Indeed, when the layers are almost independent, we inevitably have an overall even number of helical pairs that  can hybridize and get gapped out, consistently with the fact that the bulk topological invariant is defined only modulo two, so that $\nu_{\rm bi}=\nu_{\rm mono}+\nu_{\rm mono}=1+1\equiv0$~mod~2. 
Analogously, a trilayer should be non-trivial, and for thicker layers we would expect an alternation of trivial and non-trivial topology that, in the bulk limit, would give rise to a weak topological phase~\cite{Fu2007} relying on the translational invariance along the stacking direction.

In bulk jacutingaite, the layers can not be considered as nearly independent, so that this scenario is expected to break down. Indeed, first-principles simulations have shown that nearby layers are strongly hybridized, giving rise to a large second-nearest layer hopping~\cite{Marrazzo2020}. This strong coupling drives bulk jacutingaite into a semimetallic state endowed with a dual topology~\cite{Marrazzo2020,facio_prm_2019,Ghosh2020} that combines a non-zero mirror Chern number with a weak topology. 
Recent experiments have verified both the semi-metallic nature of bulk Pt\2HgSe\3~\cite{Mauro2020,Pei2021} and the presence of surface states protected by the crystalline mirror symmetry~\cite{Cucchi2020}.

\begin{figure*}
    \centering
    \includegraphics[width=\linewidth]{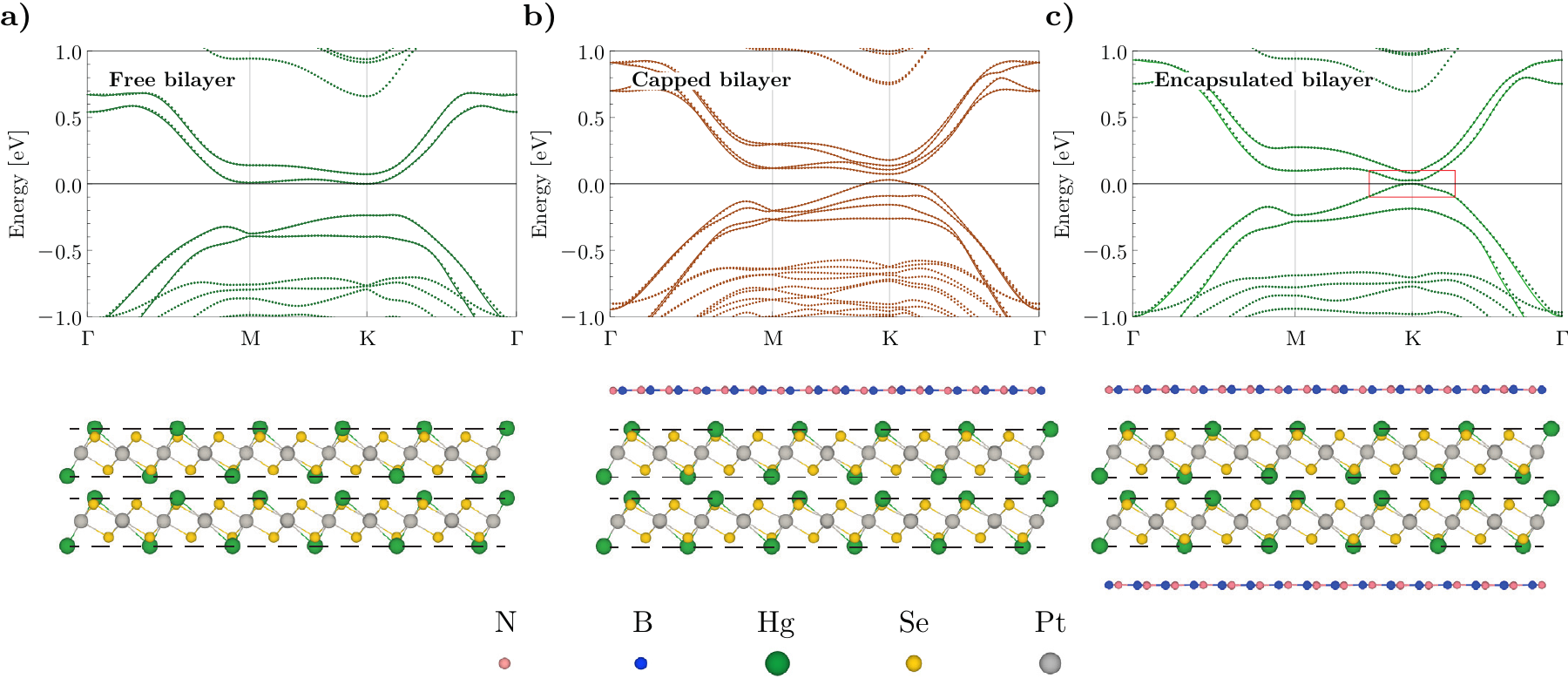}
    \caption{Band structure and lateral view of the crystal structure for: a) free, b) h-BN capped, and c) hBN-encapsulated bilayer jacutingaite.  In the top panels,  the full band structure along the path ${\bm\Gamma} - {\bf M} - {\bf K} - {\bm\Gamma}$ is shown with dots, while lines represent the Wannierized band structure of the 8 bands closest to the Fermi level. The red rectangle in c) highlights the region around the gap magnified in Fig.~\ref{fig:Gaps}. In the bottom panels, we note that in each Pt\2HgSe\3 layer Hg atoms form a buckled honeycomb lattice (see also Fig.~\ref{fig:WFStructure} for a top view in case c)), with  Hg atoms in the two sublattices alternating  above/below a Pt\2Se\3 layer. Dashed lines highlight the vertical position of Hg planes, showing the inequivalence of inner and outer planes, the latter tending to extend further away from the Pt\2Se\3 layer. This tendency, most apparent in the free bilayer, is suppressed by the presence of h-BN, with the encapsulated bilayer recovering a more symmetric structure of each layer.}
    \label{fig:CrystalBandStructures}
\end{figure*}

Here we consider bilayer jacutingaite and predict using first-principles simulations that it is trivial, although in an unexpected way, with $\nu_{\rm bi}=0+0$. The trivial gap arises from an inversion-symmetry breaking in each layer, with competing contributions from a structural distortion and the different environment affecting intra-sublattice hopping. As a result of this sublattice imbalance, the Kane-Mele term that drives the topological nature of monolayer jacutingaite\cite{Marrazzo2018} is replaced by a spin-orbit coupling that has the same sign on the two sublattices. 
When encapsulated in hexagonal boron nitride (h-BN), the trivial gap is strongly reduced and can be turned topological by a small perpendicular field, promoting bilayer Pt\2HgSe\3 into a promising system for experimental explorations.

\section{Bilayer structures}

Jacutingaite comprises AA-stacked honeycomb lattices of Hg atoms, where the A (B) sublattice is positioned above (below) a plane of Pt atoms (see Fig.~\ref{fig:CrystalBandStructures}). In the absence of spin-orbit coupling, the electronic band structure of monolayer jacutingaite contains gapless Dirac cones at the corners ${\bf K}$ and ${\bf K}'$ of the hexagonal Brillouin zone~\cite{Marrazzo2018}, similarly to what happens in graphene. These cones can be gapped in two different ways. The first is by breaking the sublattice (inversion) symmetry, e.g.\ by making the Hg distance to the Pt planes different on the two sides, leading to a trivial insulator. The second way to open a gap is via Kane-Mele spin-orbit coupling, making monolayer jacutingaite a quantum spin Hall insulator. 

In bilayer jacutingaite, although a global inversion symmetry connecting the two layers is still present, there is no inversion symmetry {\em per layer}, i.e.\ the two sublattices in each layer are no longer equivalent. This means that Hg atoms can be displaced to make each layer by itself trivial ($\nu_{\rm bi}=0+0=0$). If no or only small displacements occur and the two layers are almost independent, however, the combination of two topological monolayers together makes the bilayer trivial - following the heuristic rule $\nu_{\rm bi}=1+1=0$. In either case, bilayer jacutingaite is expected to be a trivial insulator.

\begin{table}[t]
\caption{\label{TablePos} Main properties of various layered jacutingaite  structures. Monolayer and bulk have been studied before~\cite{Marrazzo2018,Marrazzo2020}, here we show new results for free,  capped, and encapsulated bilayers, as well as for trilayer Pt\2HgSe\3 (see also App.~\ref{Sec:trilayer}). The second column contains the gap at ${\bf K}$ obtained using approximate DFT (see App.~\ref{app:methods}). The third column contains the $z$-position of the Hg atoms relative to the nearest Pt plane (see Fig.~\ref{fig:CrystalBandStructures}). Notice that in capped, encapsulated and the trilayer case the position depends on the layer (abbreviated ``l.'').}
\begin{tabular}{c|c|c}
	\hline \hline
	System 	& Gap at ${\bf K}$ & $\Delta z$  Hg-Pt \\ \hline \hline
	Bulk		& 	$<$ 1 meV	& 1.84 \AA \\ \hline
	Monolayer	& 	168 meV	& 1.73 \AA \\ \hline
	Bilayer	& 290 meV & 1.81 \AA \  (inner Hg) \\
			& 	& 2.02 \AA\ (outer Hg) \\ \hline
	BN/Bilayer	& 44 meV & 1.78 \AA\ (outer Hg, capped l.) \\
			& 	& 1.80 \AA \ (inner Hg, capped l.) \\	
			& 	& 1.78 \AA \ (inner Hg, free l.) \\
			& 	&  1.90 \AA \ (outer Hg, free l.) \\ \hline
	BN/Bilayer/BN& 27 meV	& 1.80 \AA \ (inner Hg) \\
			& 	& 1.78 \AA \ (outer Hg) \\ \hline
	Trilayer	& $<$ 1 meV & 1.84 \AA \  (middle l.) \\
			&  & 1.81 \AA \ (inner Hg, outer l.) \\
			&  & 2.02 \AA \ (outer Hg, outer l.) \\ \hline
	\hline
\end{tabular}
\end{table}

To confirm this, we perform first principles density-functional theory (DFT) calculations of various few-layered jacutingaite structures using {\sc Quantum ESPRESSO}~\cite{QE-2009,QE-2017}, with a Coulomb cut-off~\cite{Sohier2017} to reproduce the correct open boundary conditions in the vertical direction and the van-der-Waals compliant functional vdw-DF-cx~\cite{Dion2004,Berland2014,Berland2015} that gives the best agreement with Raman experiments for the vibrational frequencies of bulk jacutingaite~\cite{Mauro2020} (for further details of the calculations see App.~\ref{app:methods}). The unit cell is hexagonal (point group $D_{3d}$ or $\bar3m$) with in-plane lattice constant fixed to the bulk relaxed value $a=7.384$~\AA.

We first relax the structure of a free-standing bilayer, which shows a large shift in the vertical position of the outer Hg atoms. The distance of the Hg atoms to the Pt planes increases from 1.73 \AA\ in the monolayer to 2.02 \AA\ for the outer Hg atoms. The sublattice asymmetry is responsible for a large trivial ($\nu_{\rm bi}=0+0=0$) band-gap of 290 meV at the ${\bf K}$ point (see App.~\ref{Sec:FreeBS} and the discussion below). The same level of displacement is found in trilayer jacutingaite (see also App.~\ref{Sec:trilayer}), however, the trilayer is semimetallic owing to a large second-nearest layer hopping between the outer layers, similarly to what happens in the bulk~\cite{Marrazzo2020}. See Table~\ref{TablePos} for an overview of the Hg positions and bandgaps of the different studied structures.

The bandgap in free bilayer jacutingaite is so large that a topological transition cannot be obtained using reasonable perpendicular external electric fields (up to 1~V/\AA), unlike monolayer jacutingaite which has a transition at $E_{\rm ext}=0.36$~V/\AA. This is because the external electric field is never sufficient to reduce the sublattice asymmetry. We note in passing that these values  do not correspond to a potential drop across the system (e.g.\ $E_{\rm ext}=1$~V/\AA\ does not correspond to 1 V over 1~\AA), as in first-principles simulations we can set only the external electric field (related the dielectric displacement $D$ through $E_{\rm ext} = 4\pi D$), and not the total electric field as in experiments (through the gate voltages applied to electrodes at the two sides of the system).

A possible way to restore sublattice symmetry, and thus to make a topological transition more feasible --at least in one layer, is to suppress the lattice distortion by encapsulating one side of bilayer jacutingaite with h-BN. As can be seen in Table~\ref{TablePos} and App.~\ref{Sec:FreeBS}, this indeed reduces the sublattice asymmetry in one of the layers and reduces the gap. However, the system is still $0+0=0$ trivial and fields up to 1~V/\AA \  do not induce a topological transition - instead they make the system metallic, with a charge transfer from the bilayer to h-BN.

\begin{figure}
    \centering
    \includegraphics[width=\columnwidth]{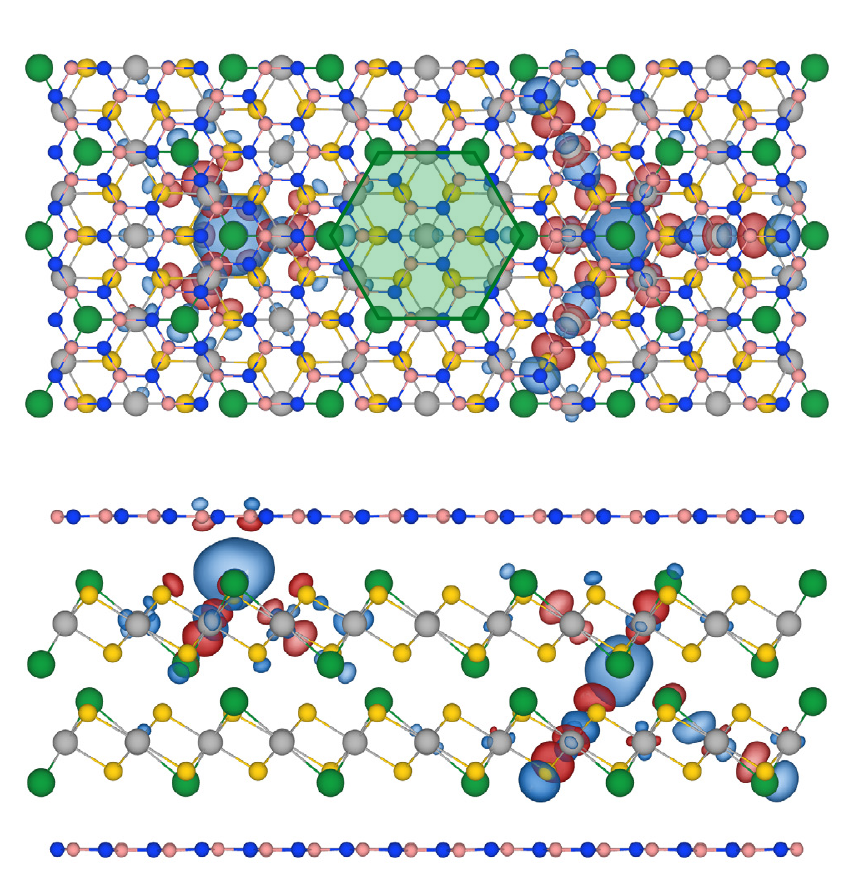}
    \caption{Top and lateral views of the crystal structure of h-BN encapsulated bilayer jacutingaite. In the top view, the green shaded area highlights the hexagonal Wigner-Seitz unit cell and the (buckled) honeycomb lattice formed by Hg atoms. Two Wannier functions associated with the two sublattices of the upper layer are also reported as isosurfaces for both positive (blue) and negative (red) values. The Wannier functions associated with the bottom layer can be simply obtained by inversion symmetry.}
    \label{fig:WFStructure}
\end{figure}

On the other hand, the reduction in gap size suggests that fully encapsulating bilayer jacutingaite with h-BN might bring us to the regime where we can induce a topological transition via an electric field. We indeed find that the Hg positions are nearly symmetrical in h-BN encapsulated bilayers. Furthermore, the gap is reduced to only 27 meV, which brings us in the regime that allows for a topological transition.

\begin{figure}
    \centering
    \includegraphics[height=0.33\columnwidth]{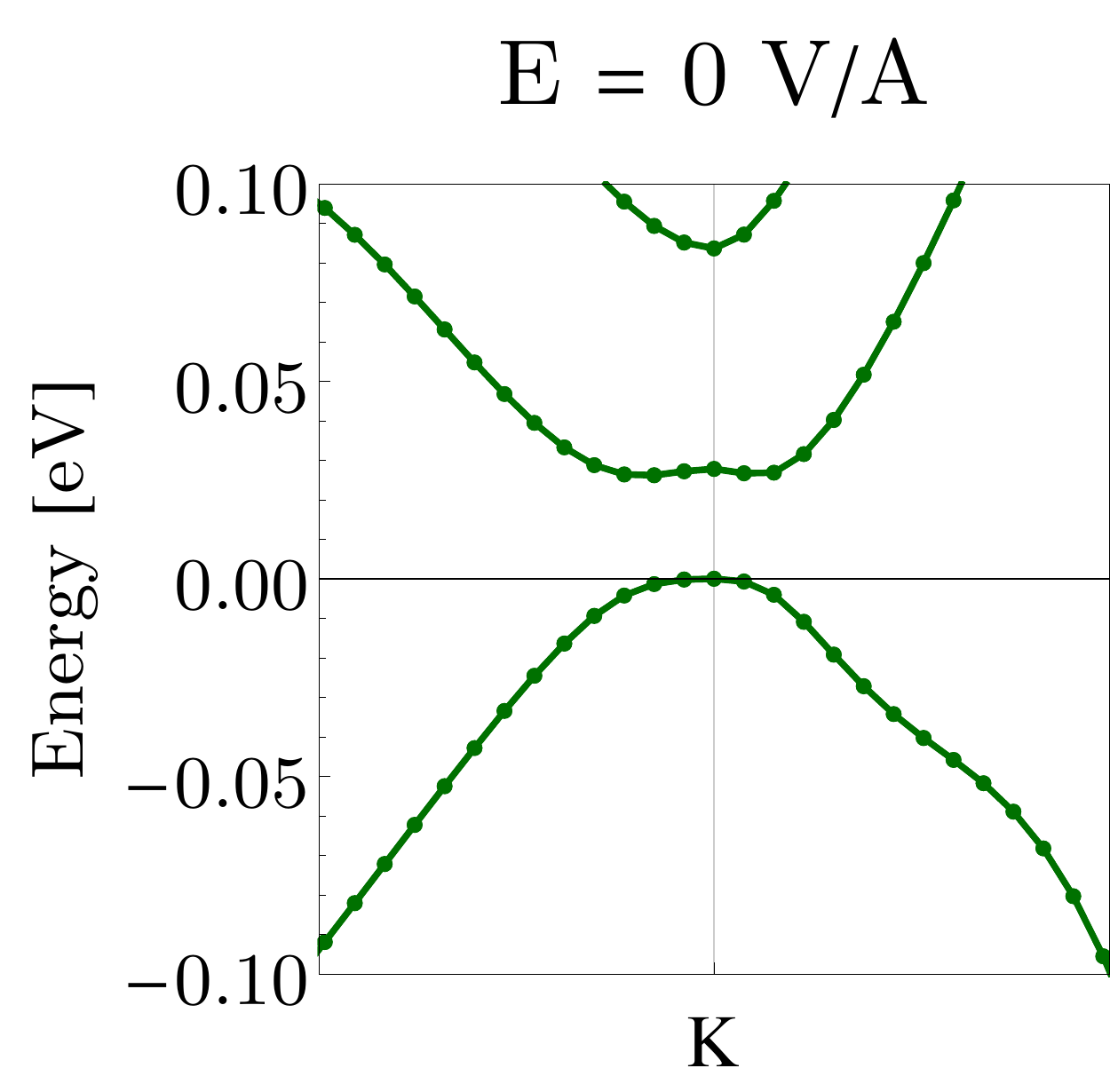}
    \includegraphics[height=0.33\columnwidth]{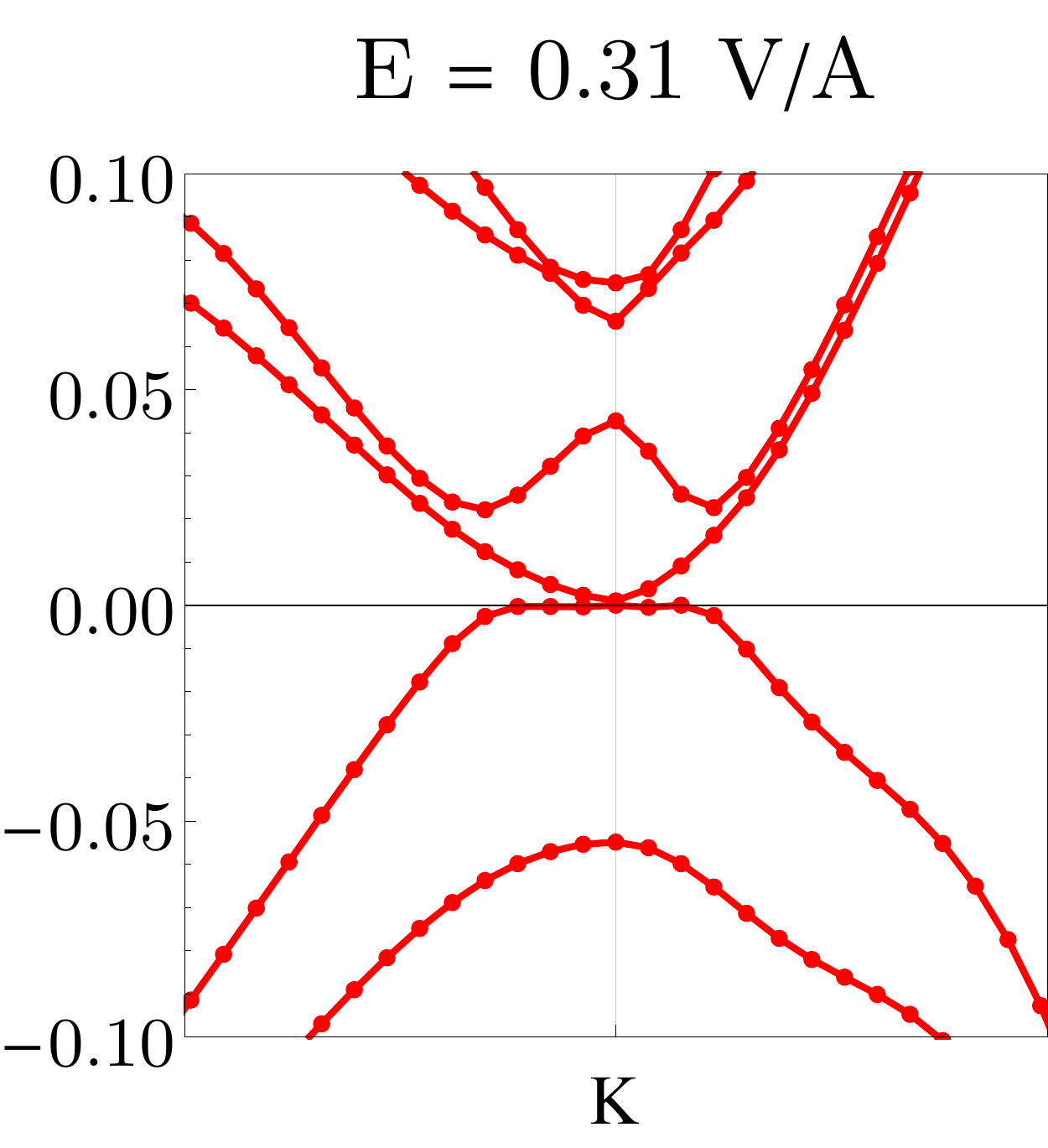}
    \includegraphics[height=0.33\columnwidth]{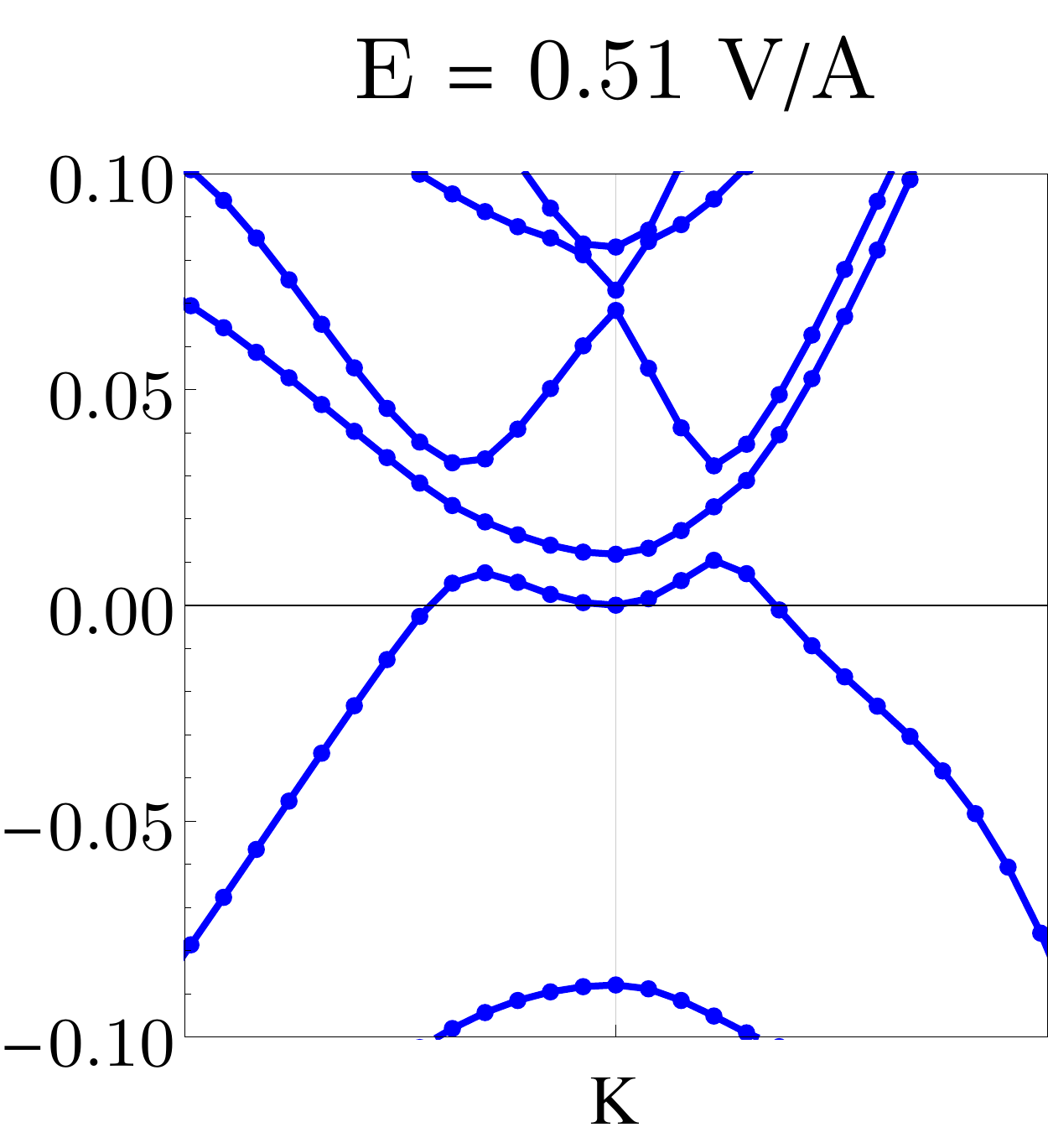}\\
    \includegraphics[width=\columnwidth]{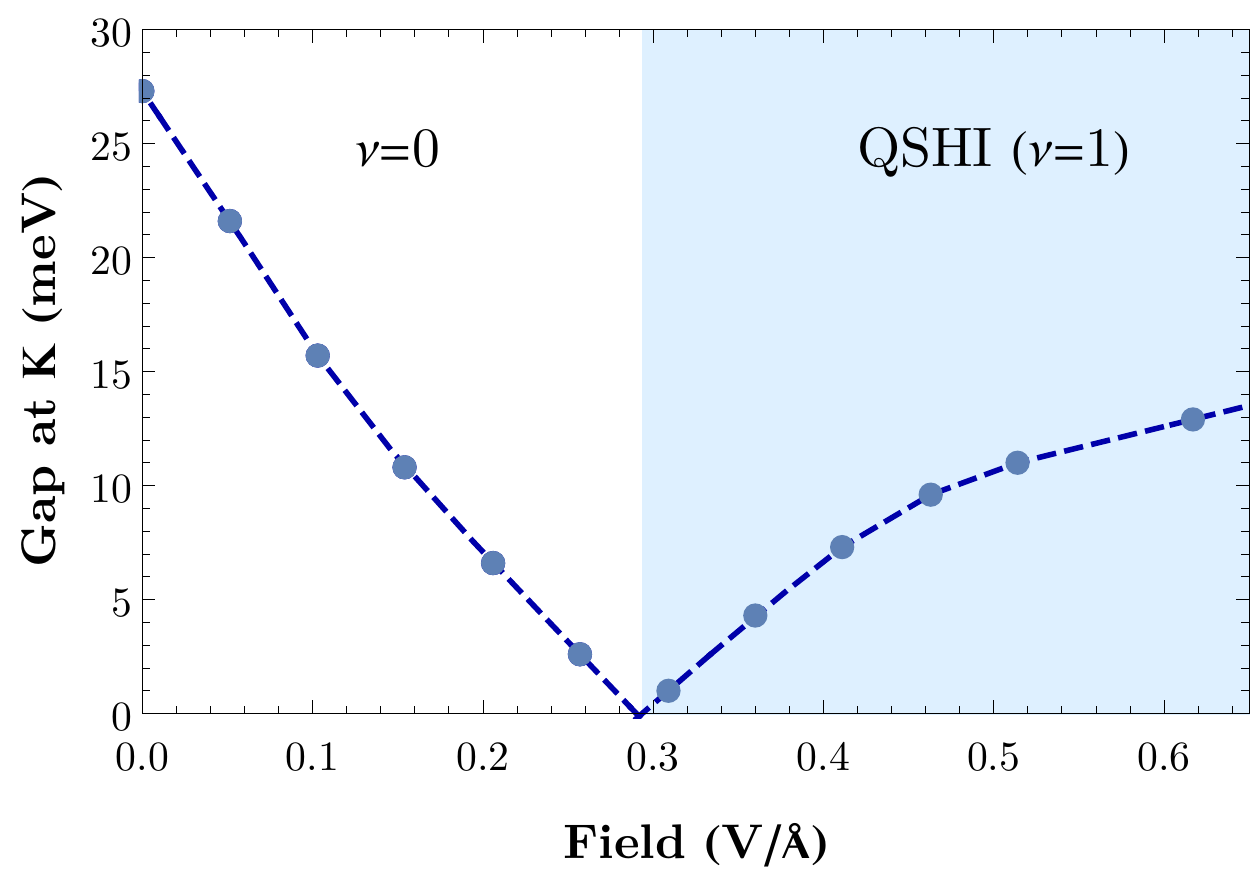}
    \caption{Top panels: Band structure of h-BN encapsulated bilayer jacutingaite around the {\bf K} point in a small energy window close the Fermi energy for three different values of the applied external electric field. Note that at zero field all bands are double degenerate due to inversion symmetry, whereas at nonzero fields this degeneracy is lifted. Bottom panel: The direct gap at the ${\bf K}$-point  as a function of perpendicular external electric field. Around $E_{\rm ext} = 0.3$ V/\AA \, the gap closes. For larger fields, a band inversion occurs and the system is a quantum spin Hall insulator with $\nu=1$. The indirect bandgap is shown in App.~\ref{Sec:IndirectGap}, while the direct gap computed with different functionals is reported in App.~\ref{Sec:HSE}.}
    \label{fig:Gaps}
\end{figure}

\section{Band structure and Wannierization}

We will now discuss in-depth the properties of the encapsulated bilayer structure. For simplicity, we locate the h-BN layers on the two sides of the bilayer so that, in the absence of an external electric field, the total system again contains inversion symmetry, and we let only the vertical position of the h-BN layers relax. The full band structure is shown in the top panel of Fig.~\ref{fig:CrystalBandStructures}c) for $E_{\rm ext}=0$, where all bands are doubly degenerate as a consequence of time-reversal and inversion symmetry. Whereas the bandwidth of the four main bands is about 1 eV, the bandgap is only 27 meV (see the top left panel of Fig.~\ref{fig:Gaps} for a closer look at the band structure around {\bf K} near the Fermi energy).

When a finite external electric field $E_{\rm ext}$ is applied, the gap at ${\bf K}$ reduces, until it closes at about $E_{\rm ext} = 0.3$~V/\AA, as shown in the top central panel of Fig.~\ref{fig:Gaps}. At larger fields, the gap reopens and increases with $E_{\rm ext}$, with the bands at ${\bf K}$ that are inverted, as shown in the right top panel of Fig.~\ref{fig:Gaps} for $E_{\rm ext}=0.5$~V/\AA. The full dependence of the direct gap size at ${\bf K}$ as a function of external electric field is reported in the bottom panel of Fig.~\ref{fig:Gaps}. It is important to stress that the calculated value of the critical field at which the topological transition occurs might depend on the choice of approximate DFT and the corresponding evolution of the energy gap with $E_{\rm ext}$. Moreover, since approximate semilocal DFT typically tends to underestimate energy gaps with respect to experiments, the critical field might be severely underestimated. To test the reliability of the above predictions we have thus performed hybrid-functional calculations, which are expected to provide more realistic estimates of the energy band gap~\cite{Borlido2019}, see App.~\ref{Sec:HSE}. The semilocal and hybrid-functional estimates give a good approximation of the lower and upper bounds of the critical field $E_{\rm ext}$, which is to be compared with experimental results.

To elucidate if the band inversion is associated with a topological phase transition, we map first-principles eigenstates for the bands facing the energy gap into a set of maximally-localized Wannier functions (WFs)~\cite{wannier_review_12}  using Wannier90~\cite{Pizzi2020}. The Hg s-orbitals are used as first projections to initialize the Wannierization procedure, as in the case of monolayer~\cite{Marrazzo2018} and bulk~\cite{Marrazzo2020} Pt\2HgSe\3. We thus end up with four WFs (eight by including spin), two per layer, that are depicted in Fig.~\ref{fig:WFStructure} for the top layer when $E_{\rm ext}=0$. While  for the external sublattice the WF is similar to the one of the monolayer~\cite{Marrazzo2018} and it is well localized on just one layer, the inner WF has significant contributions from orbitals in the opposite layer, signaling a strong hybridization between the layers similarly to what happens in bulk jacutingaite~\cite{Marrazzo2020}.  As a result, the center of the inner WFs is significantly shifted in the $z$-direction such that the center of the top WF of the bottom layer is {\em higher} than the one of the bottom WF of the top layer. 

\begin{figure}[h]
    \centering
    \includegraphics[width=\columnwidth]{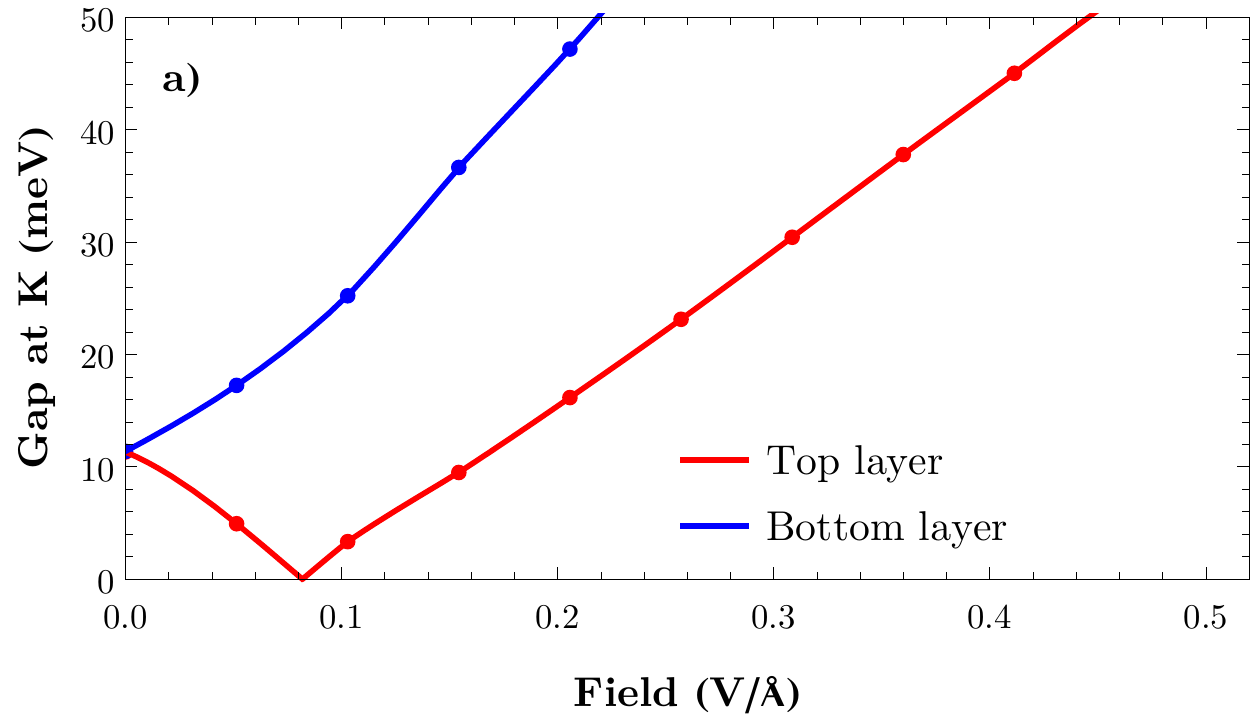}\\
    \includegraphics[width=\columnwidth]{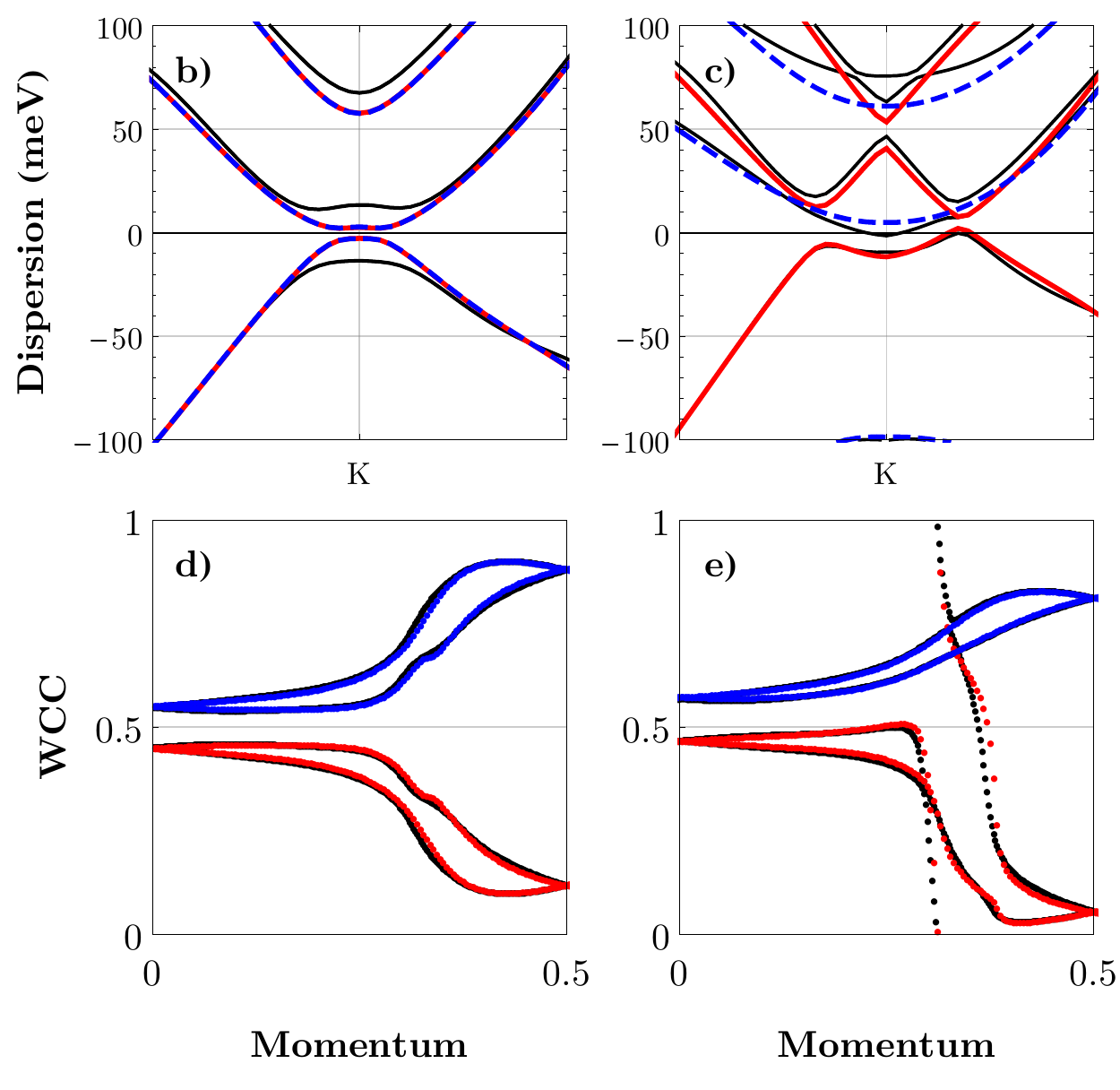}
    \caption{
    Bilayer jacutingaite can be qualitatively understood as two decoupled layers. 
    {\bf a}: To verify this, we calculated the gap at ${\bf K}$ as a function of external field for the individual layers using the tight-binding model with interlayer couplings set to zero. Though the gap is quantitatively underestimated (compare with Fig.~\ref{fig:Gaps}), we still find a topological transition in the top layer. 
    {\bf b:} The dispersion close to ${\bf K}$ changes only subtly when we have interlayer coupling (black) or not (red/blue dashed lines), at zero field. {\bf c:} At a finite external field of $E_{\mathrm{ext}} = 0.46 $~V/\AA, the dispersion for just the top layer (red) and bottom layer (blue) has still a large overlap with the full bilayer band-structure.
    {\bf d:} The WCCs (in units of the lattice parameter $a$) computed for the full tight-binding model (black) are the same as the WCCs computed per layer (red and blue). {\bf e:} Same calculation as in d, but now at a finite field $E_{\mathrm{ext}} = 0.46 $~V/\AA. We clearly see the topological nature of the bands in the top layer (red).  }
    \label{fig:DecoupledLayers}
\end{figure}

From the knowledge of the WFs, we can easily compute the $\mathbb{Z}_{2}$ topological invariant using WannierTools~\cite{wannier_tools_18} by monitoring the evolution of the Wannier charge centers (WCCs) over half of the Brillouin zone~\cite{Soluyanov2011,Yu2011}, i.e.\ the expectation value of the coordinate along one direction of hybrid WFs~\cite{Sgiarovello2001} as a function of momentum in the remaining direction, along which they are delocalized. The $\mathbb{Z}_2$ invariant $\nu$ can be then obtained by considering the parity of the number of times an arbitrary curve going from $k=0$ to $k=0.5$ (in units of the primitive reciprocal lattice vector in that direction) crosses the WCC lines~\cite{Soluyanov2011,Yu2011}. As shown in Fig.~\ref{fig:Gaps}, this confirms the presence of a topological transition as a function of the external electric field (see Fig.~\ref{fig:DecoupledLayers}d and e for the WCC evolution at small and large fields). In particular, while the system is trivial ($\nu=0$) in the absence of external fields (confirmed also using a parity approach~\cite{fu_topological_2007}), it becomes a QSHI ($\nu=1$) when $E_{\rm ext}>0.3$~V/\AA, showing that the topological state of bilayer jacutingaite can be easily manipulated. 

The mapping of the first-principles results into WFs can be beneficial also to extract an effective tight-binding model that describes the behavior of bilayer jacutingaite, helping to gain additional physical insight into the mechanisms underlying the topological transition. The resulting eight-band model (including spin) reproduces the DFT band structure around the band gap (see Fig.~\ref{fig:CrystalBandStructures}) and involves the sites of two buckled honeycomb lattices --one for each layer-- with one orbital per site and spin (given by the WFs in Fig.~\ref{fig:WFStructure}). We find that even though the WFs associated with the inner sublattices are delocalized over the two layers (see Fig.~\ref{fig:WFStructure} and the discussion above), the effective tight-binding model is still dominated by intralayer terms. 

In Table~\ref{Table:TightBinding} we summarize the most important terms of the effective tight-binding model when $E_{\rm ext}=0$. The two largest contributions are by far the intralayer nearest-neighbor (NN) hopping $t=233$ meV and the sublattice symmetry breaking on-site term $m = 135$ meV. The absence of layer inversion symmetry allows a NN (Rashba-like) spin-orbit coupling $\lambda_R$ that is vanishing in isolated monolayers.

\begin{table}[t]
\caption{\label{Table:TightBinding} Dominant tight-binding parameters of the band structure of encapsulated bilayer jacutingaite, obtained using  Wannier90. The first column indicates the type of coupling, the second column how it acts on spin ($s$), sublattice ($\sigma$) and layer ($\tau$) space. NN stands for nearest neighbor, NNN for next-nearest neighbor. Notice that the NNN Rashba and Kane-Mele spin-orbit coupling terms are highly imbalanced between the two sublattices. 
}
\begin{tabular}{c|c|c}
	\hline \hline
	Coupling           & 	Proportional to & Value [meV] \\ \hline \hline
	Onsite potential $m$   & $\sigma^z \tau^z$  & 135 \\ \hline
	NN intralayer $t$           & $\sigma^{x,y}$ &  233 \\ \hline
	NN Rashba $\lambda_R$       & $s^{x,y} \sigma^{x,y} \tau^z $&  10 \\ \hline
	NN interlayer $t_1$       & $\sigma^{x,y} \tau^{x,y}$ &  53 \\	\hline
	NNN intralayer $t_2$       & $(1-\sigma^z \tau^z)$ &  $-34$ (outer) \\
	    & $(1+\sigma^z \tau^z)$ & 32 (inner)  \\ \hline
	NNN Rashba $\lambda_R'$ & $s^{x,y} (\sigma^z - \tau^z)$ & 28  (outer only)\\ \hline
	NNN Kane-Mele $\lambda_{\mathrm{KM}}$ & $ \sigma^z s^z(1-\sigma^z \tau^z)$ & 
	    16 (outer) \\
	    & $\sigma^z  s^z(1+\sigma^z \tau^z)$ & $-2$ (inner) \\ 
	    \hline
	NNN interlayer $t_2'$ & $\tau^{x,y}$ & $-20$ \\
	\hline \hline
\end{tabular}
\end{table}

The most important contribution that couples the two layers is a NN hopping $t_1\sim50$~meV, which --together with the other relevant interlayer term $t_2'$ in Table~\ref{Table:TightBinding}-- plays a minor role on the band structure. To verify this, we calculated the band-structure using the full tight-binding model with and without interlayer coupling. The result is shown in Fig.~\ref{fig:DecoupledLayers}. In the absence of a perpendicular field, the band-structure is marginally changed: removing the interlayer coupling mainly reduces the gap at ${\bf K}$. The interlayer coupling can therefore be neglected for a qualitative understanding of the topological transition. 

An external field now reduces the gap further in the top layer whereas it increases the gap in the bottom layer (Fig.~\ref{fig:DecoupledLayers}). This is indicative of the topological transition that goes from a $\nu=0+0$ to $\nu = 0+1$ state. Indeed, Fig.~\ref{fig:DecoupledLayers}d (e) shows the evolution of the WCCs at small (large) external field. In both cases the WCCs of the full model (black) are consistent with the WCC computed for the separate layers (blue and red), thus justifying the assumption that the  topological invariant can be expressed as the sum of the invariants in the two layers, $\nu_{\rm bi} = \nu_1+\nu_2$. Moreover, while at small fields both layers are trivial (and related by inversion symmetry), at large fields the layers are no longer equivalent and the top layer (red) is non-trivial after the gap re-opens at {\bf K}.

Within each layer, the inequivalence of the two sublattices not only makes the on-site energy very different (as expressed by $m$), but also  introduces a large imbalance in the intralayer next-nearest neighbor (NNN) hopping terms. This arises as a result of the different (de)localization of the Wannier orbitals for the inner and outer sublattice sites. A first important example of NNN term is the hopping energy $t_2$ that takes approximately opposite values for the inner and outer sublattices. In particular, we find that $t_2$ is positive for the outer sites similarly to what happens in monolayer Pt\2HgSe\3, while it is negative for inner sites in complete analogy with bulk jacutingaite. This effectively \emph{staggered} NNN hopping term gives rise to a trivial gap at {\bf K} that is found to compete with the trivial gap associated with the on-site $m$ (see also below). This compensation is almost perfect in h-BN encapsulated bilayers (contrary to the free-standing case), thus explaining the very small trivial gap at {\bf K}.

Even more compelling, the imbalance affects also two additional NNN spin-orbit terms: a Kane-Mele~\cite{kane_z2_05,kane_quantum_2005} and in-plane Rashba-like~\cite{liu_buckledSOC_11} spin-orbit coupling. Also in this case, these terms retain values very close to the one in the monolayer for the outer sublattices~\cite{Marrazzo2018,Wu2019,Liu2020}, while they are strongly suppressed for the inner sublattices, in analogy with bulk jacutingaite where the effect of spin-orbit coupling is almost negligible~\cite{Marrazzo2020,facio_prm_2019,Ghosh2020}. Traditionally, topological transitions were understood in terms of spin-orbit couplings that were identical on both sublattices~\cite{kane_z2_05,kane_quantum_2005}. In bilayer jacutingaite, however, the fact that the spin-orbit coupling is different on the two sublattices requires an extension of the original Kane-Mele model.

\section{Imbalanced Kane-Mele model}

\begin{figure}[t]
    \centering
    \includegraphics[width=\columnwidth]{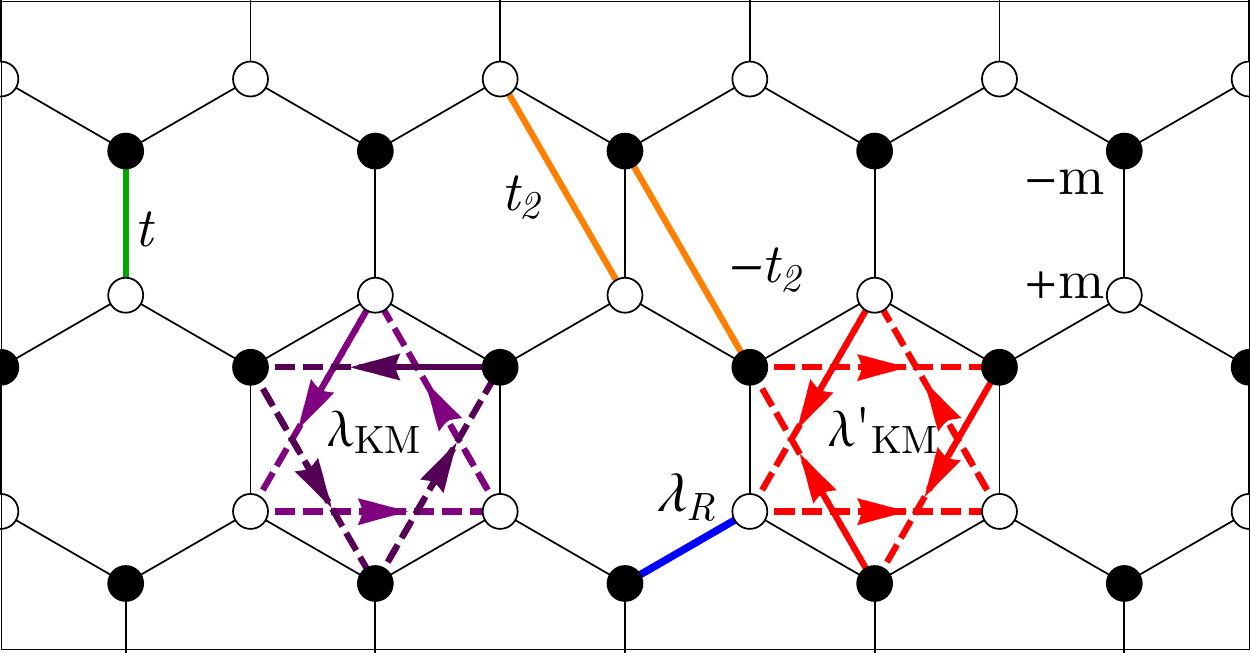}
    \caption{Terms in the imbalanced Kane-Mele model. We include nearest-neighbor $t$ and next-nearest-neighbor hopping $t_2$. Sublattice symmetry breaking is included by the term $m \sigma^z$. The spin-orbit terms include the regular Kane-Mele term $\lambda_{\mathrm{KM}} \sigma^z s^z$ and the new {\em sublattice-symmetric Kane-Mele term} $\lambda_{\mathrm{KM}}' s^z$. The arrow directions indicate the sign of the imaginary hopping. Finally, we include a nearest neighbor Rashba term $\lambda_R$ that stabilizes an unusual topological insulator phase when $\lambda_{\mathrm{KM}}' > \lambda_{\mathrm{KM}}$.}
    \label{fig:KaneMele}
\end{figure}

As argued in the previous section, the topological transition can be qualitatively understood by decoupling the two layers and focusing on the top layer only. We will now explore the question of whether we can understand the transition purely in terms of a short-ranged hopping model. To this end, we introduce the so-called {\em imbalanced Kane-Mele} model, which contains nearest ($t$) and a staggered next-nearest neighbor hopping ($t_2$), a sublattice symmetry breaking potential ($m$) and {\em two} spin-orbit terms, see  Fig.~\ref{fig:KaneMele}. In addition to the regular Kane-Mele (KM) term $
    i \lambda_{\mathrm{KM}} \sum_{\langle \langle ij \rangle \rangle} 
    \nu_{ij} c_i^\dagger s^z c_j$, which has opposite sign on the two sublattices,
we include a {\em sublattice-symmetric} Kane-Mele term $\lambda_{\mathrm{KM}}'$,
\begin{equation}
    i \lambda_{\mathrm{KM}}' \sum_{\langle \langle ij \rangle \rangle} 
    \nu_{ij} c_i^\dagger \sigma^z s^z c_j.
\end{equation}
This term, as is shown in Fig.~\ref{fig:KaneMele}, has the {\em same sign} of the spin-orbit coupling term on the two sublattices. 
As a consequence, the effective spin-orbit coupling on the two sublattices given by $\lambda_{\rm KM} \pm \lambda_{\rm KM}'$.

In momentum space, the regular KM term is proportional to $ \lambda_{\mathrm{KM}} d({\bf k})\sigma^z s^z$ where $ d({\bf k}) = 2 \sin (k_x a) - 4 \sin( k_x a/2) \cos (\sqrt{3} k_ya/2) $\cite{kane_z2_05}, if we define the honeycomb lattice with lattice vectors ${\bf a}_{1,2} = \frac{a}{2} ( \pm 1 , \sqrt{3})$. Consequently, the sublattice-symmetric KM term is proportional to $\lambda_{\mathrm{KM}}' d ({\bf k}) s^z$. As a result, at the ${\bf K}$ and ${\bf K}'$ points, the Hamiltonian reads
\begin{equation}
    H = \left(m -3t_2\right) \sigma^z + 3\sqrt{3} \lambda_{\mathrm{KM}} \kappa \sigma^z s^z  + 
     3 \sqrt{3} \lambda_{\mathrm{KM}}' \kappa s^z 
\end{equation}
where $\kappa=\pm1$ for the {\bf K}/{\bf K}$'$ valley. In the absence of spin-orbit coupling, the trivial gap at ${\bf K}$ is determined by the sublattice potential reduced by the staggered nearest neighbor hopping, $m - 3 t_2$. For $\lambda_{\mathrm{KM}}> \lambda_{\mathrm{KM}}'>0$, the gap is insensitive to the sublattice-symmetric KM term, and given by $\Delta = |m - 3 t_2| - 3 \sqrt{3} \lambda_{\mathrm{KM}} $. As long as this parameter $\Delta$ is positive, the system is trivial, and for $\Delta <0$ the model is a quantum spin Hall insulator with $\nu = 1$. 
When the two spin-orbit terms are exactly equal, $\lambda_{\mathrm{KM}} = \lambda_{\mathrm{KM}}'$, the system realizes a semimetal with quadratic band-touching as long as $\Delta <0$. If the sublattice-symmetric KM term dominates, $\lambda_{\mathrm{KM}} < \lambda_{\mathrm{KM}}'$, the system is either metallic ($\Delta' <0$) or a trivial insulator ($\Delta' >0$), with $\Delta' = |m - 3 t_2| - 3 \sqrt{3} \lambda_{\mathrm{KM}}'$. 

When also the NN Rashba spin-orbit coupling 
\begin{equation}
    i \lambda_{R} \sum_{\langle ij \rangle} c_i^\dagger ({\bf s}\times{\bf d}_{ij})^z c_j
\end{equation}
(arising from the inversion symmetry breaking in each layer) is included, not only a finite gap is opened in the semimetallic phase when $\lambda_{\mathrm{KM}} < \lambda_{\mathrm{KM}}'$, but also  a non-trivial topological state emerges for $m-3t_2<\lambda_{\rm KM}'$. When $m-3t_2$ further decreases the gap closes again (away from {\bf K}/{\bf K}$'$ at 3 Dirac cones around each corner of the Brillouin zone) and the system enters a trivial phase adiabatically connected to the one for $\Delta'>0$ when $\lambda_R=0$. The topological phase thus survives over a  finite interval of values of $m-3t_2$, whose extension increases with $\lambda_R$ and is non-vanishing only when $m-3t_2$ has the same sign as $\lambda_{\rm KM}'$, even in the limit $\lambda_{\rm KM}\to0$.

From Table~\ref{Table:TightBinding} it follows that in bilayer jacutingaite the sublattice-symmetric KM term dominates: $\lambda_{\rm KM}'=9$~meV while $\lambda_{\rm KM}=7$~meV (their sum is the 'outer' sublattice Kane-Mele term while their difference the inner). In the absence of a field, the sublattice potential $m$ controls the physics and we expect a trivial insulator. A perpendicular electric field affects the value of $m$ and $t_2$, because of the different $z$-positions of the Wannier orbitals. In particular, by wannierizing the electronic structure at different $E_{\rm ext}$, we find that the onsite potential and the staggered hopping $t_2$ change linearly with field, approximately as $\Delta m = -$ 135 meV per V/\AA\ and $\Delta t_2 = 9$~meV per V/\AA. This change has opposite sign in the two layers, making the bottom layer having a larger trivial gap upon the application of a field, whereas the top layer reduces the gap.

Consistently with the results above, the imbalanced Kane-Mele thus predicts that the bottom layer remains trivial with just an increasing gap at {\bf K}, while in the top layer the gap decreases and closes at a critical value when the field is such that  $m-3t_2=\lambda_{\rm KM}'$, in qualitative agreement with Fig.~\ref{fig:DecoupledLayers}a). When $E_{\rm ext}$ is further increased, the gap re-opens and the top layer is in a topologically non-trivial state ($\nu=1$), protected by the NN Rashba $\lambda_R$. This prediction is validated by the fact that without spin-flipping inter-sublattice hopping terms (such as $\lambda_R$), no topological transition occurs even in the full WF tight-binding model. Of course, longer-range and interlayer hopping terms in the full model play a role in the quantitative determination of energy gaps and transition fields, but the imbalanced Kane-Mele is sufficient to describe the essential physical features of the topological transition occurring in encapsulated bilayer jacutingaite.

\section{Outlook}

We predict that h-BN encapsulated bilayer Pt\2HgSe\3 undergoes a topological transition under the application of an electric field, from a trivial insulator at zero field to a quantum spin Hall insulator. 
The transition can be qualitatively understood by considering the layers as decoupled and described by an imbalanced Kane-Mele model, with a new, sublattice-symmetric next-nearest-neighbor spin-orbit coupling. This additional term emerges from the inversion-symmetry breaking in each layer associated with the inequivalence of the two sublattices. This imbalance also allows for a non-zero Rashba spin-orbit coupling that plays an essential role in stabilizing the topological phase at large fields.

Jacutingaite has been predicted to be potentially exfoliable~\cite{Marrazzo2018,Mounet2018}, and consequently bilayer jacutingaite can also appear during an exfoliation process. Recently, experiments have shown that this is possible~\cite{Kandrai2020}, although the quality of the exfoliated samples needs to be improved. If encapsulated, this could lead to the construction of a gate-switchable topological insulator (off at zero field, on at finite field), which is complementary to the monolayer case (on at zero field, off at finite field)~\cite{Marrazzo2018}.

\section{Acknowledgements}
We greatly acknowledge Antimo Marrazzo for fruitful discussions. Support has been provided by the Swiss National Science Foundation (SNSF) through the Ambizione program (M.G.\ grant PZ00P2\_174056, L.R.\ grant 	 PZ00P2\_174208). M.G.\ acknowledges support also from the Italian Ministry for University and Research through the Levi-Montalcini program and from the NCCR MARVEL funded by the SNSF. Simulation time was provided by CSCS on Piz Daint (production project s917).

\appendix
\renewcommand\thesection{A\arabic{section}}
\renewcommand\theequation{A\arabic{equation}}
\renewcommand\thefigure{A\arabic{figure}}
\setcounter{equation}{0}
\setcounter{figure}{0}
\setcounter{table}{0}
\setcounter{section}{0}

\begin{figure}[t]
    \centering
    \includegraphics[width=\columnwidth]{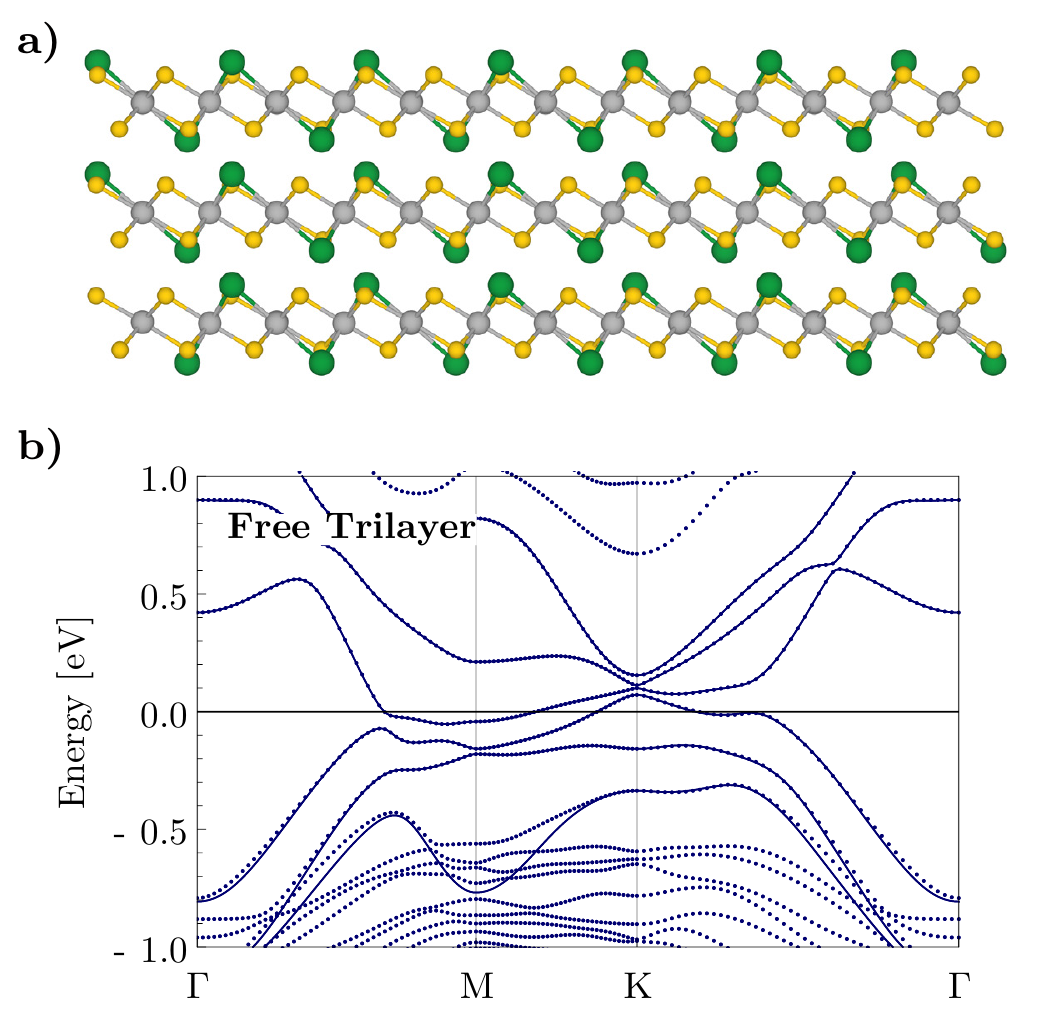} 

    \caption{a) Lateral view of the crystal structure of free trilayer jacutingaite. The central layer is symmetric, while in the other layers the outermost Hg atoms are further away from the central plane of Pt atoms with respect to the inner ones. b) Electronic band structure of  trilayer jacutingaite, where symbols represent direct calculations while lines are the result of a minimal tight-binding model based on Wannier functions. The zero of energy is set at the Fermi level.}
    \label{fig:trilayer}
\end{figure}

\begin{figure}[t]
    \centering
    \includegraphics[width=\columnwidth]{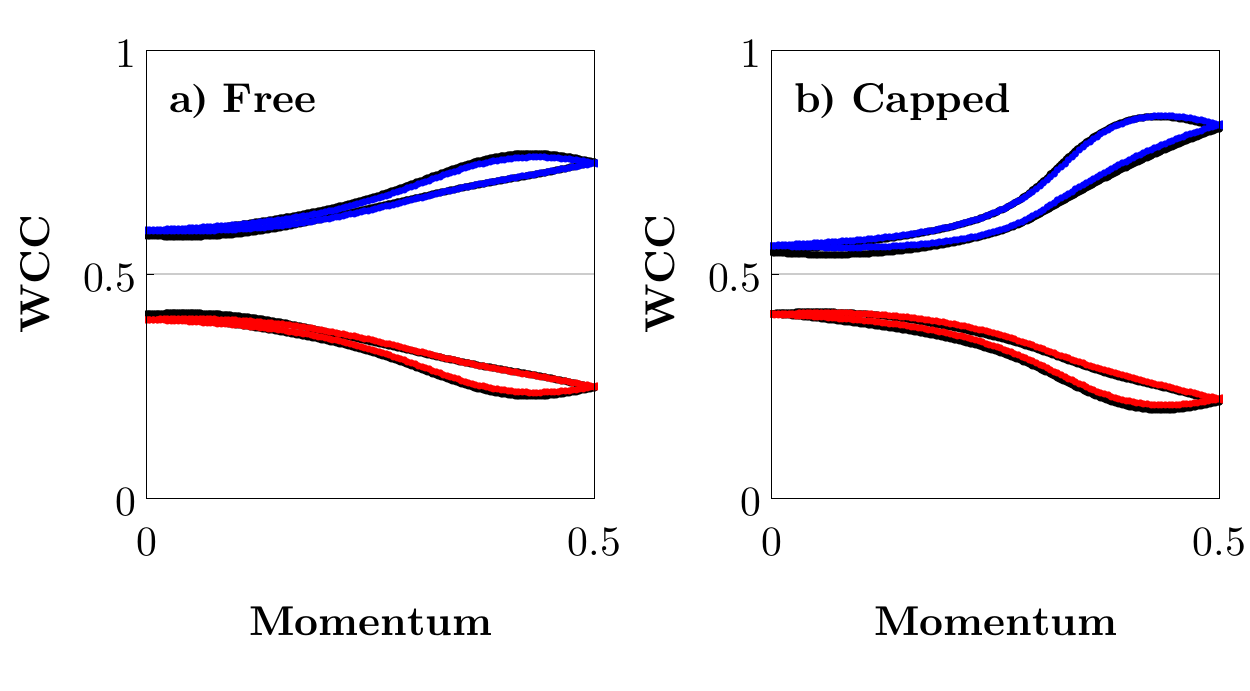} 

    \caption{Evolution of Wannier charge centers (WCCs) for (a) free and (b) h-BN capped bilayer Pt\2HgSe\3.
    Both systems are topologically trivial.  Note that the WCCs (in units of the lattice parameter $a$) computed for the full tight-binding model (black) are the same as the WCCs computed per layer (red and blue).}
    \label{fig:FreeCappedWCC}
\end{figure}

\begin{figure}[t]
    \centering
    \includegraphics[width=\columnwidth]{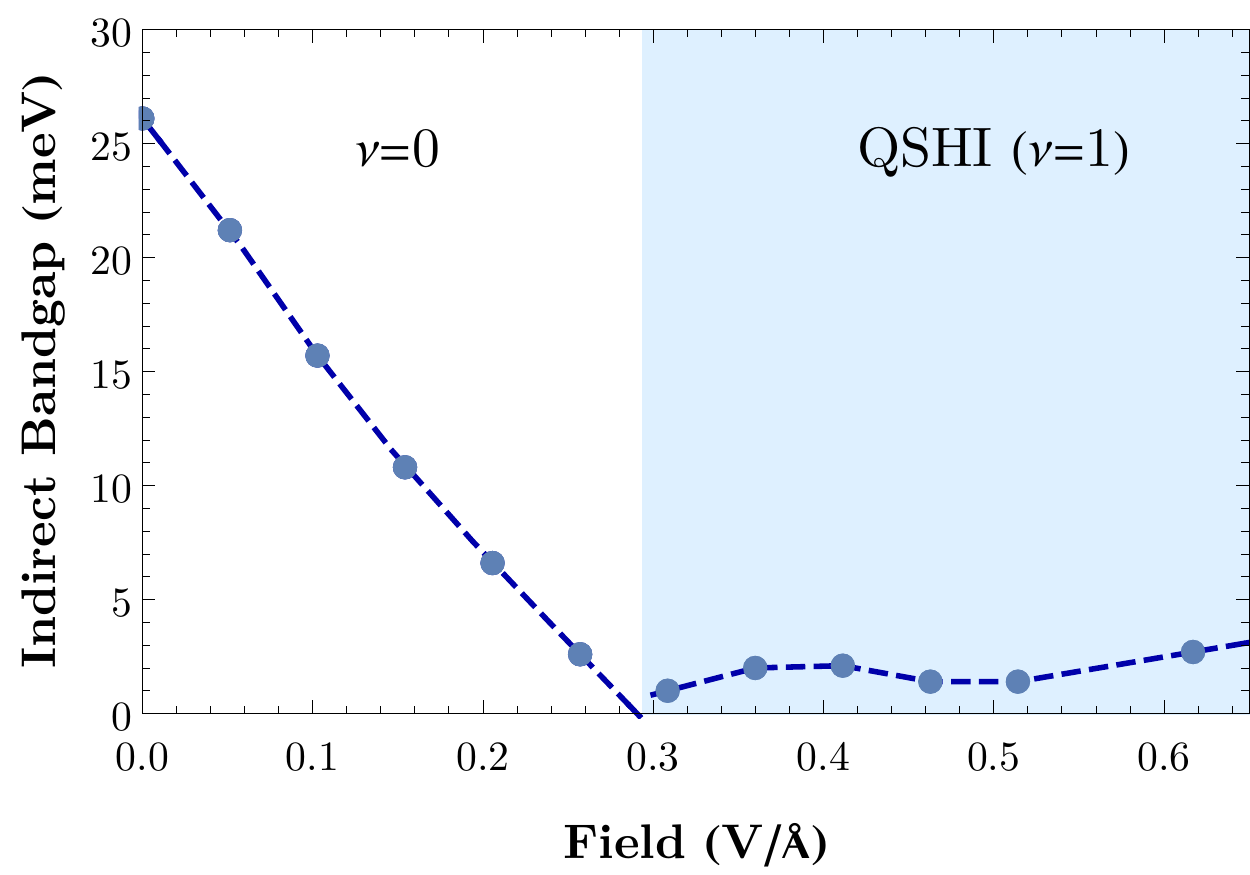}
    \caption{The indirect band gap (measured as minimum of the conduction band minus the maximum of the valence band) of the encapsulated bilayer as a function of external electric field. In the topological phase the gap is smaller than the direct gap at {\bf K} in Fig.~\ref{fig:Gaps} but still positive, suggesting the presence of a fully-developed band gap.}
    \label{fig:IndirectGap}
\end{figure}

\begin{figure}[t]
    \centering
    \includegraphics[width=\columnwidth]{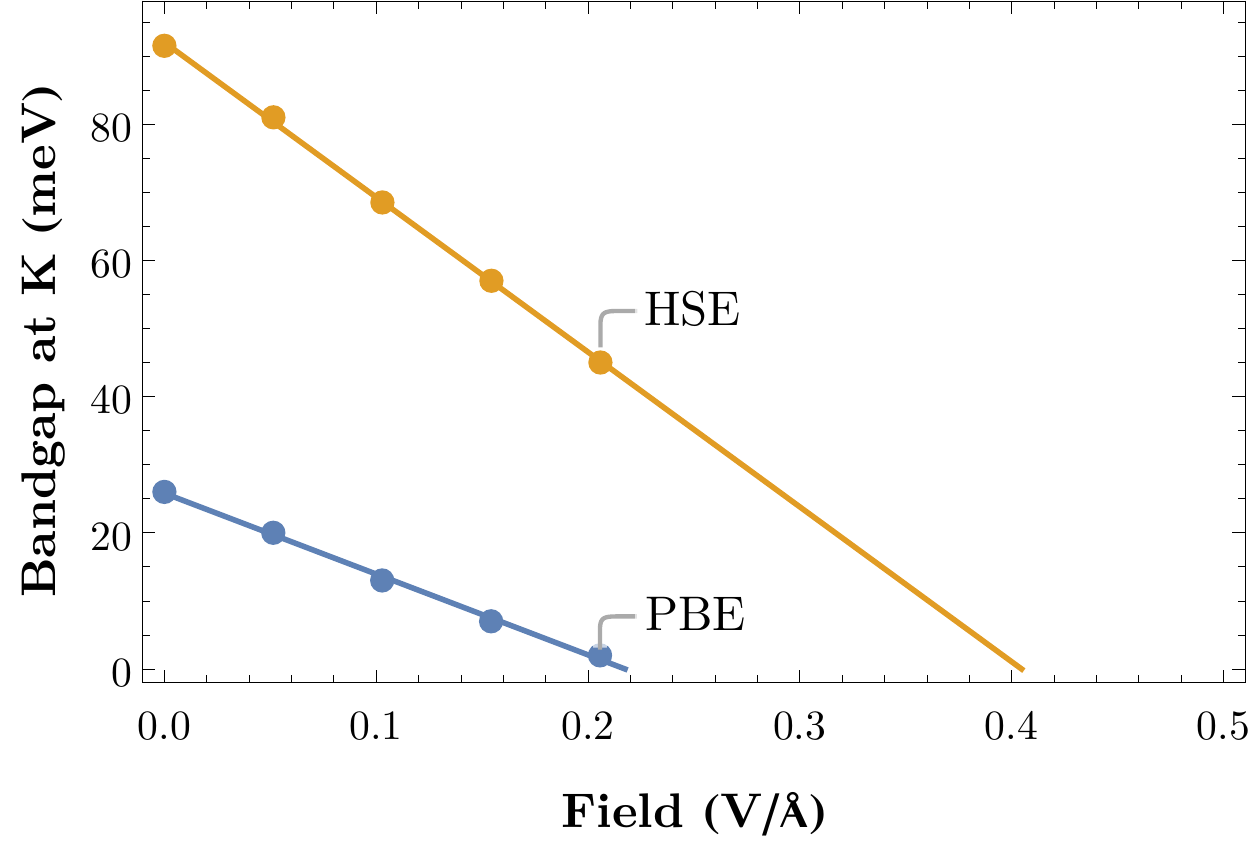}
    \caption{Direct band gap at the {\bf K} point for  hBN-encapsulated bilayer jacutingaite as a function of the external electric field computed using either the PBE~\cite{perdew_pbe_96} or the HSE~\cite{HSE} functional. Dots represent actual calculation results, while lines are linear extrapolations.}
    \label{fig:HSE}
\end{figure}

\section{Calculation details}\label{app:methods}

As mentioned in the main text, all first-principles DFT calculations were performed using the {\sc Quantum ESPRESSO} suite of codes~\cite{QE-2009,QE-2017}. Structural relaxations were carried out using the ``cx" variant~\cite{Berland2014} of the van-der-Waals compliant vdw-DF~\cite{Dion2004,Berland2015} functional without spin-orbit coupling, with pseudopotentials from the Standard Solid State Pseudopotential library~\cite{prandini_precision_2018} (efficiency version 1.0), with an energy cutoff of 60 Ry for wavefunctions and 480 Ry for the density. The Brillouin zone was sampled with $8\times8\times1$ k-points of a uniform $\Gamma$-centered Monkhorst-Pack grid with a cold smearing of 0.015~Ry~\cite{mv_smearing_99}. Band structures were computed by including spin-orbit coupling through fully-relativistic pseudopotentials of the Pseudo-Dojo family~\cite{dojo_paper_18} with a wavefunction cutoff of 80~Ry on top of self-consistent calculations with $12\times12\times1$ k-points within the generalized gradient approximation as formulated by Perdew-Burke-Ernzerhof (PBE)~\cite{perdew_pbe_96}. Calculations with the HSE hybrid functional~\cite{HSE} have been performed with norm-conserving pseudopotentials\cite{hamann_oncv_13} from the SG15 library~\cite{Schlipf2015,Scherpelz2016} that do not have non-linear core corrections, using a cutoff of 50~Ry both for wavefunctions and the representation of the Fock operator, and a $6\times6$ k-point grid.

Crystal structures and Wannier functions are visualized using VESTA~\cite{VESTA}.

\section{Trilayer crystal and band structure}
\label{Sec:trilayer}
Although the main target is  bilayer jacutingaite, we have also considered the trilayer structure. 
In Fig.~\ref{fig:trilayer} we show both the relaxed crystal structure and the computed electronic band structure along a high-symmetry path in the Brillouin zone. 
We note that contrary to the bilayer cases shown in Fig.~\ref{fig:CrystalBandStructures}, the system is metallic, mainly due to a strong interlayer coupling between the outermost layers similar  to the second-nearest-layer hopping of bulk jacutingaite~\cite{Marrazzo2020}.

\section{Wannier charge centers for free and capped bilayer}
\label{Sec:FreeBS}

In the main text we introduced in addition to the encapsulated bilayer jacutingaite also a free bilayer and a h-BN capped bilayer (with h-BN only on one side), whose band structure is shown in Fig.~\ref{fig:CrystalBandStructures}. In Fig.~\ref{fig:FreeCappedWCC} we plot evolution of the Wannier charge centers, to show that the free and capped bilayers are  topologically trivial. In particular, the Wannier charge centers computed assuming the layers to be decoupled (red and blue) are the same as for the full system (black), suggesting that the topological invariant for the bilayer can be expressed as the sum of the invariants of the two layers, $\nu_{\rm bi} = \nu_1 + \nu_2$, with both $\nu_1=\nu_2 = 0$.

\section{Indirect gap in encapsulated bilayer}
\label{Sec:IndirectGap}

In the topological phase, the direct gap at ${\bf K}$ of the encapsulated bilayer in Fig.~\ref{fig:Gaps} is not equal to the full band gap. This is typical for band-inversion, and is visible in the band structure of Fig.~\ref{fig:Gaps}. Nevertheless, the maximum of the valence band remains below the minimum of the conduction band, i.e.\ there is a fully developed band gap, whose magnitude is shown in Fig.~\ref{fig:IndirectGap}.

\section{Direct gap with hybrid functional}
\label{Sec:HSE}

Standard approximations to DFT, including the generalized-gradient PBE approximation~\cite{perdew_pbe_96} used here (see App.~\ref{app:methods}), tend to largely underestimate the energy gap, so that topological transitions and the corresponding critical electric field might also be affected. To test the reliability of the conclusions in the main text, we report here results for the direct gap at ${\bf K}$ (which controls the topological transition) of the encapsulated bilayer using hybrid functionals (in particular the HSE functional~\cite{HSE}), which typically lead to estimates of the energy gap in closer agreement with experiments~\cite{Borlido2019}. 

Fig.~\ref{fig:HSE} shows that  for $E_{\rm ext}=0$ the gap is largely underestimated by almost a factor of 4 in PBE-DFT with respect to hybrid-functional calculations. Still, the rate at which the gap closes as a function of the external electric field is much larger with the HSE functional than with PBE (note that the latter results slightly differ from Fig.~\ref{fig:Gaps} because smearing is not used in this case and thus the almost linear behavior extends down to zero gap while deviations associated with smearing appear in Fig.~\ref{fig:Gaps}). As a consequence, a topological phase transition still occurs even at the HSE level and the estimated critical field is only a factor of 2 larger than in PBE calculations, supporting the robustness of the phenomena discussed in the main text.


\begin{thebibliography}{10}%
\makeatletter
\providecommand \@ifxundefined [1]{%
 \ifx #1\undefined \expandafter \@firstoftwo
 \else \expandafter \@secondoftwo
\fi
}%
\providecommand \@ifnum [1]{%
 \ifnum #1\expandafter \@firstoftwo
 \else \expandafter \@secondoftwo
\fi
}%
\providecommand \enquote [1]{``#1''}%
\providecommand \bibnamefont  [1]{#1}%
\providecommand \bibfnamefont [1]{#1}%
\providecommand \citenamefont [1]{#1}%
\providecommand\href[0]{\@sanitize\@href}%
\providecommand\@href[1]{\endgroup\@@startlink{#1}\endgroup\@@href}%
\providecommand\@@href[1]{#1\@@endlink}%
\providecommand \@sanitize [0]{\begingroup\catcode`\&12\catcode`\#12\relax}%
\@ifxundefined \pdfoutput {\@firstoftwo}{%
 \@ifnum{\z@=\pdfoutput}{\@firstoftwo}{\@secondoftwo}%
}{%
 \providecommand\@@startlink[1]{\leavevmode}%
 \providecommand\@@endlink[0]{}%
}{%
 \providecommand\@@startlink[1]{%
  \leavevmode
  \pdfstartlink
   attr{/Border[0 0 1 ]/H/I/C[0 1 1]}%
   user{/Subtype/Link/A<</Type/Action/S/URI/URI(#1)>>}%
  \relax
 }%
 \providecommand\@@endlink[0]{\pdfendlink}%
}%
\providecommand \url  [0]{\begingroup\@sanitize \@url }%
\providecommand \@url [1]{\endgroup\@href {#1}{\urlprefix}}%
\providecommand \urlprefix [0]{URL }%
\providecommand \Eprint[0]{\href }%
\@ifxundefined \urlstyle {%
  \providecommand \doi [1]{doi:\discretionary{}{}{}#1}%
}{%
  \providecommand \doi [0]{doi:\discretionary{}{}{}\begingroup
  \urlstyle{rm}\Url }%
}%
\providecommand \doibase [0]{http://dx.doi.org/}%
\providecommand \Doi[1]{\href{\doibase#1}}%
\providecommand \bibAnnote [3]{%
  \BibitemShut{#1}%
  \begin{quotation}\noindent
    \textsc{Key:}\ #2\\\textsc{Annotation:}\ #3%
  \end{quotation}%
}%
\providecommand \bibAnnoteFile [2]{%
  \IfFileExists{#2}{\bibAnnote {#1} {#2} {\input{#2}}}{}%
}%
\providecommand \typeout [0]{\immediate \write \m@ne }%
\providecommand \selectlanguage [0]{\@gobble}%
\providecommand \bibinfo [0]{\@secondoftwo}%
\providecommand \bibfield [0]{\@secondoftwo}%
\providecommand \translation [1]{[#1]}%
\providecommand \BibitemOpen[0]{}%
\providecommand \bibitemStop [0]{}%
\providecommand \bibitemNoStop [0]{.\EOS\space}%
\providecommand \EOS [0]{\spacefactor3000\relax}%
\providecommand \BibitemShut [1]{\csname bibitem#1\endcsname}%
\bibitem{review_hasankane_2010}%
  \BibitemOpen
  \bibfield{author}{%
  \bibinfo {author} {\bibfnamefont{M.~Z.}\ \bibnamefont{Hasan}}\ and\ \bibinfo
  {author} {\bibfnamefont{C.~L.}\ \bibnamefont{Kane}},\ }%
  \Doi{10.1103/RevModPhys.82.3045}{\emph{\bibinfo {title} {Colloquium:
  Topological insulators}}},\ \bibinfo {journal} {Rev. Mod. Phys.}\
  \textbf{\bibinfo {volume} {82}},\ \bibinfo {pages} {3045} (\bibinfo {year}
  {2010}).~%
  \bibAnnoteFile{Stop}{review_hasankane_2010}%
\bibitem{bernevig_topological_2013}%
  \BibitemOpen
  \bibfield{author}{%
  \bibinfo {author} {\bibfnamefont{B.~A.}\ \bibnamefont{Bernevig}}\ and\
  \bibinfo {author} {\bibfnamefont{T.~L.}\ \bibnamefont{Hughes}},\ }%
  \emph{\bibinfo {title} {Topological insulators and topological
  superconductors}}\ (\bibinfo {publisher} {Princeton University Press},\
  \bibinfo {year} {2013}).~%
  \bibAnnoteFile{Stop}{bernevig_topological_2013}%
\bibitem{Kitaev2009}%
  \BibitemOpen
  \bibfield{author}{%
  \bibinfo {author} {\bibfnamefont{A.}~\bibnamefont{Kitaev}},\ }%
  \Doi{10.1063/1.3149495}{\emph{\bibinfo {title} {Periodic table for
  topological insulators and superconductors}}},\ \bibinfo {journal} {AIP
  Conference Proceedings}\ \textbf{\bibinfo {volume} {1134}},\ \bibinfo {pages}
  {22} (\bibinfo {year} {2009}).~%
  \bibAnnoteFile{Stop}{Kitaev2009}%
\bibitem{Ryu2010}%
  \BibitemOpen
  \bibfield{author}{%
  \bibinfo {author} {\bibfnamefont{S.}~\bibnamefont{Ryu}}, \bibinfo {author}
  {\bibfnamefont{A.~P.}\ \bibnamefont{Schnyder}}, \bibinfo {author}
  {\bibfnamefont{A.}~\bibnamefont{Furusaki}},\ and\ \bibinfo {author}
  {\bibfnamefont{A.~W.~W.}\ \bibnamefont{Ludwig}},\ }%
  \Doi{10.1088/1367-2630/12/6/065010}{\emph{\bibinfo {title} {Topological
  insulators and superconductors: tenfold way and dimensional hierarchy}}},\
  \bibinfo {journal} {New Journal of Physics}\ \textbf{\bibinfo {volume}
  {12}},\ \bibinfo {pages} {065010} (\bibinfo {year} {2010}).~%
  \bibAnnoteFile{Stop}{Ryu2010}%
\bibitem{Fu2011}%
  \BibitemOpen
  \bibfield{author}{%
  \bibinfo {author} {\bibfnamefont{L.}~\bibnamefont{Fu}},\ }%
  \Doi{10.1103/PhysRevLett.106.106802}{\emph{\bibinfo {title} {Topological
  Crystalline Insulators}}},\ \bibinfo {journal} {Phys. Rev. Lett.}\
  \textbf{\bibinfo {volume} {106}},\ \bibinfo {pages} {106802} (\bibinfo {year}
  {2011}).~%
  \bibAnnoteFile{Stop}{Fu2011}%
\bibitem{Slager2013}%
  \BibitemOpen
  \bibfield{author}{%
  \bibinfo {author} {\bibfnamefont{R.-J.}\ \bibnamefont{Slager}}, \bibinfo
  {author} {\bibfnamefont{A.}~\bibnamefont{Mesaros}}, \bibinfo {author}
  {\bibfnamefont{V.}~\bibnamefont{Juri{\v c}i{\'c}}},\ and\ \bibinfo {author}
  {\bibfnamefont{J.}~\bibnamefont{Zaanen}},\ }%
  \Doi{10.1038/nphys2513}{\emph{\bibinfo {title} {The space group
  classification of topological band-insulators}}},\ \bibinfo {journal} {Nature
  Physics}\ \textbf{\bibinfo {volume} {9}},\ \bibinfo {pages} {98} (\bibinfo
  {year} {2013}).~%
  \bibAnnoteFile{Stop}{Slager2013}%
\bibitem{TKNN}%
  \BibitemOpen
  \bibfield{author}{%
  \bibinfo {author} {\bibfnamefont{D.~J.}\ \bibnamefont{Thouless}}, \bibinfo
  {author} {\bibfnamefont{M.}~\bibnamefont{Kohmoto}}, \bibinfo {author}
  {\bibfnamefont{M.~P.}\ \bibnamefont{Nightingale}},\ and\ \bibinfo {author}
  {\bibfnamefont{M.}~\bibnamefont{den Nijs}},\ }%
  \Doi{10.1103/PhysRevLett.49.405}{\emph{\bibinfo {title} {Quantized Hall
  Conductance in a Two-Dimensional Periodic Potential}}},\ \bibinfo {journal}
  {Phys. Rev. Lett.}\ \textbf{\bibinfo {volume} {49}},\ \bibinfo {pages} {405}
  (\bibinfo {year} {1982}).~%
  \bibAnnoteFile{Stop}{TKNN}%
\bibitem{Kohmoto1985}%
  \BibitemOpen
  \bibfield{author}{%
  \bibinfo {author} {\bibfnamefont{M.}~\bibnamefont{Kohmoto}},\ }%
  \Doi{https://doi.org/10.1016/0003-4916(85)90148-4}{\emph{\bibinfo {title}
  {Topological invariant and the quantization of the Hall conductance}}},\
  \bibinfo {journal} {Annals of Physics}\ \textbf{\bibinfo {volume} {160}},\
  \bibinfo {pages} {343 } (\bibinfo {year} {1985}).~%
  \bibAnnoteFile{Stop}{Kohmoto1985}%
\bibitem{kane_quantum_2005}%
  \BibitemOpen
  \bibfield{author}{%
  \bibinfo {author} {\bibfnamefont{C.~L.}\ \bibnamefont{Kane}}\ and\ \bibinfo
  {author} {\bibfnamefont{E.~J.}\ \bibnamefont{Mele}},\ }%
  \Doi{10.1103/PhysRevLett.95.226801}{\emph{\bibinfo {title} {Quantum {{Spin
  Hall Effect}} in {{Graphene}}}}},\ \bibinfo {journal} {Physical Review
  Letters}\ \textbf{\bibinfo {volume} {95}},\ \bibinfo {pages} {226801}
  (\bibinfo {year} {2005}).~%
  \bibAnnoteFile{Stop}{kane_quantum_2005}%
\bibitem{kane_z2_05}%
  \BibitemOpen
  \bibfield{author}{%
  \bibinfo {author} {\bibfnamefont{C.~L.}\ \bibnamefont{Kane}}\ and\ \bibinfo
  {author} {\bibfnamefont{E.~J.}\ \bibnamefont{Mele}},\ }%
  \Doi{10.1103/PhysRevLett.95.146802}{\emph{\bibinfo {title} {$\mathbb{Z}_2$
  Topological Order and the Quantum Spin Hall Effect}}},\ \bibinfo {journal}
  {Physical Review Letters}\ \textbf{\bibinfo {volume} {95}},\ \bibinfo {pages}
  {146802} (\bibinfo {year} {2005}).~%
  \bibAnnoteFile{Stop}{kane_z2_05}%
\bibitem{Bernevig2006}%
  \BibitemOpen
  \bibfield{author}{%
  \bibinfo {author} {\bibfnamefont{B.~A.}\ \bibnamefont{Bernevig}}\ and\
  \bibinfo {author} {\bibfnamefont{S.-C.}\ \bibnamefont{Zhang}},\ }%
  \Doi{10.1103/PhysRevLett.96.106802}{\emph{\bibinfo {title} {Quantum Spin Hall
  Effect}}},\ \bibinfo {journal} {Phys. Rev. Lett.}\ \textbf{\bibinfo {volume}
  {96}},\ \bibinfo {pages} {106802} (\bibinfo {year} {2006}).~%
  \bibAnnoteFile{Stop}{Bernevig2006}%
\bibitem{bernevig_quantum_2006}%
  \BibitemOpen
  \bibfield{author}{%
  \bibinfo {author} {\bibfnamefont{B.~A.}\ \bibnamefont{Bernevig}}, \bibinfo
  {author} {\bibfnamefont{T.~L.}\ \bibnamefont{Hughes}},\ and\ \bibinfo
  {author} {\bibfnamefont{S.-C.}\ \bibnamefont{Zhang}},\ }%
  \Doi{10.1126/science.1133734}{\emph{\bibinfo {title} {Quantum {Spin} {Hall}
  {Effect} and {Topological} {Phase} {Transition} in {HgTe} {Quantum}
  {Wells}}}},\ \bibinfo {journal} {Science}\ \textbf{\bibinfo {volume} {314}},\
  \bibinfo {pages} {1757} (\bibinfo {year} {2006}).~%
  \bibAnnoteFile{Stop}{bernevig_quantum_2006}%
\bibitem{Konig2007}%
  \BibitemOpen
  \bibfield{author}{%
  \bibinfo {author} {\bibfnamefont{M.}~\bibnamefont{K{\"o}nig}}, \bibinfo
  {author} {\bibfnamefont{S.}~\bibnamefont{Wiedmann}}, \bibinfo {author}
  {\bibfnamefont{C.}~\bibnamefont{Br{\"u}ne}}, \bibinfo {author}
  {\bibfnamefont{A.}~\bibnamefont{Roth}}, \bibinfo {author}
  {\bibfnamefont{H.}~\bibnamefont{Buhmann}}, \bibinfo {author}
  {\bibfnamefont{L.~W.}\ \bibnamefont{Molenkamp}}, \bibinfo {author}
  {\bibfnamefont{X.-L.}\ \bibnamefont{Qi}},\ and\ \bibinfo {author}
  {\bibfnamefont{S.-C.}\ \bibnamefont{Zhang}},\ }%
  \Doi{10.1126/science.1148047}{\emph{\bibinfo {title} {Quantum Spin Hall
  Insulator State in HgTe Quantum Wells}}},\ \bibinfo {journal} {Science}\
  \textbf{\bibinfo {volume} {318}},\ \bibinfo {pages} {766} (\bibinfo {year}
  {2007}).~%
  \bibAnnoteFile{Stop}{Konig2007}%
\bibitem{Roth2009}%
  \BibitemOpen
  \bibfield{author}{%
  \bibinfo {author} {\bibfnamefont{A.}~\bibnamefont{Roth}}, \bibinfo {author}
  {\bibfnamefont{C.}~\bibnamefont{Br{\"u}ne}}, \bibinfo {author}
  {\bibfnamefont{H.}~\bibnamefont{Buhmann}}, \bibinfo {author}
  {\bibfnamefont{L.~W.}\ \bibnamefont{Molenkamp}}, \bibinfo {author}
  {\bibfnamefont{J.}~\bibnamefont{Maciejko}}, \bibinfo {author}
  {\bibfnamefont{X.-L.}\ \bibnamefont{Qi}},\ and\ \bibinfo {author}
  {\bibfnamefont{S.-C.}\ \bibnamefont{Zhang}},\ }%
  \Doi{10.1126/science.1174736}{\emph{\bibinfo {title} {Nonlocal Transport in
  the Quantum Spin Hall State}}},\ \bibinfo {journal} {Science}\
  \textbf{\bibinfo {volume} {325}},\ \bibinfo {pages} {294} (\bibinfo {year}
  {2009}).~%
  \bibAnnoteFile{Stop}{Roth2009}%
\bibitem{Grabecki2013}%
  \BibitemOpen
  \bibfield{author}{%
  \bibinfo {author} {\bibfnamefont{G.}~\bibnamefont{Grabecki}}, \bibinfo
  {author} {\bibfnamefont{J.}~\bibnamefont{Wr\'obel}}, \bibinfo {author}
  {\bibfnamefont{M.}~\bibnamefont{Czapkiewicz}}, \bibinfo {author}
  {\bibfnamefont{L.}~\bibnamefont{Cywi\ifmmode~\acute{n}\else \'{n}\fi{}ski}},
  \bibinfo {author} {\bibfnamefont{S.}~\bibnamefont{Giera\l{}towska}}, \bibinfo
  {author} {\bibfnamefont{E.}~\bibnamefont{Guziewicz}}, \bibinfo {author}
  {\bibfnamefont{M.}~\bibnamefont{Zholudev}}, \bibinfo {author}
  {\bibfnamefont{V.}~\bibnamefont{Gavrilenko}}, \bibinfo {author}
  {\bibfnamefont{N.~N.}\ \bibnamefont{Mikhailov}}, \bibinfo {author}
  {\bibfnamefont{S.~A.}\ \bibnamefont{Dvoretski}}, \bibinfo {author}
  {\bibfnamefont{F.}~\bibnamefont{Teppe}}, \bibinfo {author}
  {\bibfnamefont{W.}~\bibnamefont{Knap}},\ and\ \bibinfo {author}
  {\bibfnamefont{T.}~\bibnamefont{Dietl}},\ }%
  \Doi{10.1103/PhysRevB.88.165309}{\emph{\bibinfo {title} {Nonlocal resistance
  and its fluctuations in microstructures of band-inverted HgTe/(Hg,Cd)Te
  quantum wells}}},\ \bibinfo {journal} {Phys. Rev. B}\ \textbf{\bibinfo
  {volume} {88}},\ \bibinfo {pages} {165309} (\bibinfo {year} {2013}).~%
  \bibAnnoteFile{Stop}{Grabecki2013}%
\bibitem{Konig2013}%
  \BibitemOpen
  \bibfield{author}{%
  \bibinfo {author} {\bibfnamefont{M.}~\bibnamefont{K\"onig}}, \bibinfo
  {author} {\bibfnamefont{M.}~\bibnamefont{Baenninger}}, \bibinfo {author}
  {\bibfnamefont{A.~G.~F.}\ \bibnamefont{Garcia}}, \bibinfo {author}
  {\bibfnamefont{N.}~\bibnamefont{Harjee}}, \bibinfo {author}
  {\bibfnamefont{B.~L.}\ \bibnamefont{Pruitt}}, \bibinfo {author}
  {\bibfnamefont{C.}~\bibnamefont{Ames}}, \bibinfo {author}
  {\bibfnamefont{P.}~\bibnamefont{Leubner}}, \bibinfo {author}
  {\bibfnamefont{C.}~\bibnamefont{Br\"une}}, \bibinfo {author}
  {\bibfnamefont{H.}~\bibnamefont{Buhmann}}, \bibinfo {author}
  {\bibfnamefont{L.~W.}\ \bibnamefont{Molenkamp}},\ and\ \bibinfo {author}
  {\bibfnamefont{D.}~\bibnamefont{Goldhaber-Gordon}},\ }%
  \Doi{10.1103/PhysRevX.3.021003}{\emph{\bibinfo {title} {Spatially Resolved
  Study of Backscattering in the Quantum Spin Hall State}}},\ \bibinfo
  {journal} {Phys. Rev. X}\ \textbf{\bibinfo {volume} {3}},\ \bibinfo {pages}
  {021003} (\bibinfo {year} {2013}).~%
  \bibAnnoteFile{Stop}{Konig2013}%
\bibitem{Knez2011}%
  \BibitemOpen
  \bibfield{author}{%
  \bibinfo {author} {\bibfnamefont{I.}~\bibnamefont{Knez}}, \bibinfo {author}
  {\bibfnamefont{R.~R.}\ \bibnamefont{Du}},\ and\ \bibinfo {author}
  {\bibfnamefont{G.}~\bibnamefont{Sullivan}},\ }%
  \Doi{10.1103/PhysRevLett.107.136603}{\emph{\bibinfo {title} {{Evidence for
  helical edge modes in inverted InAs/GaSb quantum wells}}}},\ \bibinfo
  {journal} {Physical Review Letters}\ \textbf{\bibinfo {volume} {107}},\
  \bibinfo {pages} {136603} (\bibinfo {year} {2011}).~%
  \bibAnnoteFile{Stop}{Knez2011}%
\bibitem{Suzuki2013}%
  \BibitemOpen
  \bibfield{author}{%
  \bibinfo {author} {\bibfnamefont{K.}~\bibnamefont{Suzuki}}, \bibinfo {author}
  {\bibfnamefont{Y.}~\bibnamefont{Harada}}, \bibinfo {author}
  {\bibfnamefont{K.}~\bibnamefont{Onomitsu}},\ and\ \bibinfo {author}
  {\bibfnamefont{K.}~\bibnamefont{Muraki}},\ }%
  \Doi{10.1103/PhysRevB.87.235311}{\emph{\bibinfo {title} {{Edge channel
  transport in the InAs/GaSb topological insulating phase}}}},\ \bibinfo
  {journal} {Physical Review B}\ \textbf{\bibinfo {volume} {87}},\ \bibinfo
  {pages} {235311} (\bibinfo {year} {2013}).~%
  \bibAnnoteFile{Stop}{Suzuki2013}%
\bibitem{Fei2017}%
  \BibitemOpen
  \bibfield{author}{%
  \bibinfo {author} {\bibfnamefont{Z.}~\bibnamefont{Fei}}, \bibinfo {author}
  {\bibfnamefont{T.}~\bibnamefont{Palomaki}}, \bibinfo {author}
  {\bibfnamefont{S.}~\bibnamefont{Wu}}, \bibinfo {author}
  {\bibfnamefont{W.}~\bibnamefont{Zhao}}, \bibinfo {author}
  {\bibfnamefont{X.}~\bibnamefont{Cai}}, \bibinfo {author}
  {\bibfnamefont{B.}~\bibnamefont{Sun}}, \bibinfo {author}
  {\bibfnamefont{P.}~\bibnamefont{Nguyen}}, \bibinfo {author}
  {\bibfnamefont{J.}~\bibnamefont{Finney}}, \bibinfo {author}
  {\bibfnamefont{X.}~\bibnamefont{Xu}},\ and\ \bibinfo {author}
  {\bibfnamefont{D.~H.}\ \bibnamefont{Cobden}},\ }%
  \Doi{10.1038/nphys4091}{\emph{\bibinfo {title} {{Edge conduction in monolayer
  $\mathrm{WTe}_2$}}}},\ \bibinfo {journal} {Nature Physics}\ \textbf{\bibinfo
  {volume} {13}},\ \bibinfo {pages} {677} (\bibinfo {year} {2017}).~%
  \bibAnnoteFile{Stop}{Fei2017}%
\bibitem{Tang2017}%
  \BibitemOpen
  \bibfield{author}{%
  \bibinfo {author} {\bibfnamefont{S.}~\bibnamefont{Tang}}, \bibinfo {author}
  {\bibfnamefont{C.}~\bibnamefont{Zhang}}, \bibinfo {author}
  {\bibfnamefont{D.}~\bibnamefont{Wong}}, \bibinfo {author}
  {\bibfnamefont{Z.}~\bibnamefont{Pedramrazi}}, \bibinfo {author}
  {\bibfnamefont{H.~Z.}\ \bibnamefont{Tsai}}, \bibinfo {author}
  {\bibfnamefont{C.}~\bibnamefont{Jia}}, \bibinfo {author}
  {\bibfnamefont{B.}~\bibnamefont{Moritz}}, \bibinfo {author}
  {\bibfnamefont{M.}~\bibnamefont{Claassen}}, \bibinfo {author}
  {\bibfnamefont{H.}~\bibnamefont{Ryu}}, \bibinfo {author}
  {\bibfnamefont{S.}~\bibnamefont{Kahn}}, \bibinfo {author}
  {\bibfnamefont{J.}~\bibnamefont{Jiang}}, \bibinfo {author}
  {\bibfnamefont{H.}~\bibnamefont{Yan}}, \bibinfo {author}
  {\bibfnamefont{M.}~\bibnamefont{Hashimoto}}, \bibinfo {author}
  {\bibfnamefont{D.}~\bibnamefont{Lu}}, \bibinfo {author}
  {\bibfnamefont{R.~G.}\ \bibnamefont{Moore}}, \bibinfo {author}
  {\bibfnamefont{C.~C.}\ \bibnamefont{Hwang}}, \bibinfo {author}
  {\bibfnamefont{C.}~\bibnamefont{Hwang}}, \bibinfo {author}
  {\bibfnamefont{Z.}~\bibnamefont{Hussain}}, \bibinfo {author}
  {\bibfnamefont{Y.}~\bibnamefont{Chen}}, \bibinfo {author}
  {\bibfnamefont{M.~M.}\ \bibnamefont{Ugeda}}, \bibinfo {author}
  {\bibfnamefont{Z.}~\bibnamefont{Liu}}, \bibinfo {author}
  {\bibfnamefont{X.}~\bibnamefont{Xie}}, \bibinfo {author}
  {\bibfnamefont{T.~P.}\ \bibnamefont{Devereaux}}, \bibinfo {author}
  {\bibfnamefont{M.~F.}\ \bibnamefont{Crommie}}, \bibinfo {author}
  {\bibfnamefont{S.~K.}\ \bibnamefont{Mo}},\ and\ \bibinfo {author}
  {\bibfnamefont{Z.~X.}\ \bibnamefont{Shen}},\ }%
  \Doi{10.1038/nphys4174}{\emph{\bibinfo {title} {{Quantum spin Hall state in
  monolayer $\mathrm{1T'-WTe}_2$}}}},\ \bibinfo {journal} {Nature Physics}\
  \textbf{\bibinfo {volume} {13}},\ \bibinfo {pages} {683} (\bibinfo {year}
  {2017}).~%
  \bibAnnoteFile{Stop}{Tang2017}%
\bibitem{Wu2018}%
  \BibitemOpen
  \bibfield{author}{%
  \bibinfo {author} {\bibfnamefont{S.}~\bibnamefont{Wu}}, \bibinfo {author}
  {\bibfnamefont{V.}~\bibnamefont{Fatemi}}, \bibinfo {author}
  {\bibfnamefont{Q.~D.}\ \bibnamefont{Gibson}}, \bibinfo {author}
  {\bibfnamefont{K.}~\bibnamefont{Watanabe}}, \bibinfo {author}
  {\bibfnamefont{T.}~\bibnamefont{Taniguchi}}, \bibinfo {author}
  {\bibfnamefont{R.~J.}\ \bibnamefont{Cava}},\ and\ \bibinfo {author}
  {\bibfnamefont{P.}~\bibnamefont{Jarillo-Herrero}},\ }%
  \Doi{10.1126/science.aan6003}{\emph{\bibinfo {title} {{Observation of the
  quantum spin Hall effect up to 100 kelvin in a monolayer crystal}}}},\
  \bibinfo {journal} {Science}\ \textbf{\bibinfo {volume} {359}},\ \bibinfo
  {pages} {76} (\bibinfo {year} {2018}).~%
  \bibAnnoteFile{Stop}{Wu2018}%
\bibitem{Shi2019}%
  \BibitemOpen
  \bibfield{author}{%
  \bibinfo {author} {\bibfnamefont{Y.}~\bibnamefont{Shi}}, \bibinfo {author}
  {\bibfnamefont{J.}~\bibnamefont{Kahn}}, \bibinfo {author}
  {\bibfnamefont{B.}~\bibnamefont{Niu}}, \bibinfo {author}
  {\bibfnamefont{Z.}~\bibnamefont{Fei}}, \bibinfo {author}
  {\bibfnamefont{B.}~\bibnamefont{Sun}}, \bibinfo {author}
  {\bibfnamefont{X.}~\bibnamefont{Cai}}, \bibinfo {author}
  {\bibfnamefont{B.~A.}\ \bibnamefont{Francisco}}, \bibinfo {author}
  {\bibfnamefont{D.}~\bibnamefont{Wu}}, \bibinfo {author}
  {\bibfnamefont{Z.~X.}\ \bibnamefont{Shen}}, \bibinfo {author}
  {\bibfnamefont{X.}~\bibnamefont{Xu}}, \bibinfo {author}
  {\bibfnamefont{D.~H.}\ \bibnamefont{Cobden}},\ and\ \bibinfo {author}
  {\bibfnamefont{Y.~T.}\ \bibnamefont{Cui}},\ }%
  \Doi{10.1126/sciadv.aat8799}{\emph{\bibinfo {title} {{Imaging quantum spin
  Hall edges in monolayer $\mathrm{WTe}_2$}}}},\ \bibinfo {journal} {Science
  Advances}\ \textbf{\bibinfo {volume} {5}},\ \bibinfo {pages} {1} (\bibinfo
  {year} {2019}).~%
  \bibAnnoteFile{Stop}{Shi2019}%
\bibitem{Marrazzo2018}%
  \BibitemOpen
  \bibfield{author}{%
  \bibinfo {author} {\bibfnamefont{A.}~\bibnamefont{Marrazzo}}, \bibinfo
  {author} {\bibfnamefont{M.}~\bibnamefont{Gibertini}}, \bibinfo {author}
  {\bibfnamefont{D.}~\bibnamefont{Campi}}, \bibinfo {author}
  {\bibfnamefont{N.}~\bibnamefont{Mounet}},\ and\ \bibinfo {author}
  {\bibfnamefont{N.}~\bibnamefont{Marzari}},\ }%
  \Doi{10.1103/PhysRevLett.120.117701}{\emph{\bibinfo {title} {Prediction of a
  Large-Gap and Switchable {Kane-Mele} Quantum Spin {Hall} Insulator}}},\
  \bibinfo {journal} {Phys. Rev. Lett.}\ \textbf{\bibinfo {volume} {120}},\
  \bibinfo {pages} {117701} (\bibinfo {year} {2018}).~%
  \bibAnnoteFile{Stop}{Marrazzo2018}%
\bibitem{Luo2021}%
  \BibitemOpen
  \bibfield{author}{%
  \bibinfo {author} {\bibfnamefont{F.}~\bibnamefont{Luo}}, \bibinfo {author}
  {\bibfnamefont{X.}~\bibnamefont{Hao}}, \bibinfo {author}
  {\bibfnamefont{Y.}~\bibnamefont{Jia}}, \bibinfo {author}
  {\bibfnamefont{J.}~\bibnamefont{Yao}}, \bibinfo {author}
  {\bibfnamefont{Q.}~\bibnamefont{Meng}}, \bibinfo {author}
  {\bibfnamefont{S.}~\bibnamefont{Zhai}}, \bibinfo {author}
  {\bibfnamefont{J.}~\bibnamefont{Wu}}, \bibinfo {author}
  {\bibfnamefont{W.}~\bibnamefont{Dou}},\ and\ \bibinfo {author}
  {\bibfnamefont{M.}~\bibnamefont{Zhou}},\ }%
  \Doi{10.1039/D0NR06889F}{\emph{\bibinfo {title} {Functionalization induced
  quantum spin Hall to quantum anomalous Hall phase transition in monolayer
  jacutingaite}}},\ \bibinfo {journal} {Nanoscale}\ \textbf{\bibinfo {volume}
  {13}},\ \bibinfo {pages} {2527} (\bibinfo {year} {2021}).~%
  \bibAnnoteFile{Stop}{Luo2021}%
\bibitem{Liu2020}%
  \BibitemOpen
  \bibfield{author}{%
  \bibinfo {author} {\bibfnamefont{Z.}~\bibnamefont{Liu}}, \bibinfo {author}
  {\bibfnamefont{Y.}~\bibnamefont{Han}}, \bibinfo {author}
  {\bibfnamefont{Y.}~\bibnamefont{Ren}}, \bibinfo {author}
  {\bibfnamefont{Q.}~\bibnamefont{Niu}},\ and\ \bibinfo {author}
  {\bibfnamefont{Z.}~\bibnamefont{Qiao}},\ }%
  \href{https://arxiv.org/abs/2012.13298}{\emph{\bibinfo {title} {Van der Waals
  Heterostructure $\mathrm{Pt_2HgSe_3/CrI_3}$ for Topological
  Valleytronics}}},\ \bibinfo {journal} {arXiv.org:2012.13298}\  (\bibinfo
  {year} {2020}).~%
  \bibAnnoteFile{Stop}{Liu2020}%
\bibitem{Mounet2018}%
  \BibitemOpen
  \bibfield{author}{%
  \bibinfo {author} {\bibfnamefont{N.}~\bibnamefont{Mounet}}, \bibinfo {author}
  {\bibfnamefont{M.}~\bibnamefont{Gibertini}}, \bibinfo {author}
  {\bibfnamefont{P.}~\bibnamefont{Schwaller}}, \bibinfo {author}
  {\bibfnamefont{D.}~\bibnamefont{Campi}}, \bibinfo {author}
  {\bibfnamefont{A.}~\bibnamefont{Merkys}}, \bibinfo {author}
  {\bibfnamefont{A.}~\bibnamefont{Marrazzo}}, \bibinfo {author}
  {\bibfnamefont{T.}~\bibnamefont{Sohier}}, \bibinfo {author}
  {\bibfnamefont{I.~E.}\ \bibnamefont{Castelli}}, \bibinfo {author}
  {\bibfnamefont{A.}~\bibnamefont{Cepellotti}}, \bibinfo {author}
  {\bibfnamefont{G.}~\bibnamefont{Pizzi}},\ and\ \bibinfo {author}
  {\bibfnamefont{N.}~\bibnamefont{Marzari}},\ }%
  \Doi{10.1038/s41565-017-0035-5}{\emph{\bibinfo {title} {Two-Dimensional
  Materials from High-Throughput Computational Exfoliation of Experimentally
  Known Compounds}}},\ \bibinfo {journal} {Nature Nanotechnology}\
  \textbf{\bibinfo {volume} {13}},\ \bibinfo {pages} {246} (\bibinfo {year}
  {2018}).~%
  \bibAnnoteFile{Stop}{Mounet2018}%
\bibitem{cabral_first_obs_08}%
  \BibitemOpen
  \bibfield{author}{%
  \bibinfo {author} {\bibfnamefont{A.~R.}\ \bibnamefont{Cabral}}, \bibinfo
  {author} {\bibfnamefont{H.~F.}\ \bibnamefont{Galbiatti}}, \bibinfo {author}
  {\bibfnamefont{R.}~\bibnamefont{Kwitko-Ribeiro}},\ and\ \bibinfo {author}
  {\bibfnamefont{B.}~\bibnamefont{Lehmann}},\ }%
  \Doi{10.1111/j.1365-3121.2007.00783.x}{\emph{\bibinfo {title} {Platinum
  Enrichment at Low Temperatures and Related Microstructures, with Examples of
  Hongshiite ({{PtCu}}) and Empirical `$\mathrm{Pt}_2\mathrm{HgSe}_3$' from
  {Itabira}, {Minas Gerais}, {Brazil}}}},\ \bibinfo {journal} {Terra Nova}\
  \textbf{\bibinfo {volume} {20}},\ \bibinfo {pages} {32} (\bibinfo {year}
  {2008}).~%
  \bibAnnoteFile{Stop}{cabral_first_obs_08}%
\bibitem{jacutingaite_exp_12}%
  \BibitemOpen
  \bibfield{author}{%
  \bibinfo {author} {\bibfnamefont{A.}~\bibnamefont{Vymazalov{\'a}}}, \bibinfo
  {author} {\bibfnamefont{F.}~\bibnamefont{Laufek}}, \bibinfo {author}
  {\bibfnamefont{M.}~\bibnamefont{Dr{\'a}bek}}, \bibinfo {author}
  {\bibfnamefont{A.~R.}\ \bibnamefont{Cabral}}, \bibinfo {author}
  {\bibfnamefont{J.}~\bibnamefont{Haloda}}, \bibinfo {author}
  {\bibfnamefont{T.}~\bibnamefont{Sidorinov{\'a}}}, \bibinfo {author}
  {\bibfnamefont{B.}~\bibnamefont{Lehmann}}, \bibinfo {author}
  {\bibfnamefont{H.~F.}\ \bibnamefont{Galbiatti}},\ and\ \bibinfo {author}
  {\bibfnamefont{J.}~\bibnamefont{Drahokoupil}},\ }%
  \Doi{10.3749/canmin.50.2.431}{\emph{\bibinfo {title} {Jacutingaite,
  $\mathrm{Pt}_2\mathrm{HgSe}_3$, a new platinum-group mineral species from the
  Cau{\^e} iron-ore deposit, Itabira district, Minas Gerais, Brazil}}},\
  \bibinfo {journal} {The Canadian Mineralogist}\ \textbf{\bibinfo {volume}
  {50}},\ \bibinfo {pages} {431} (\bibinfo {year} {2012}).~%
  \bibAnnoteFile{Stop}{jacutingaite_exp_12}%
\bibitem{Kandrai2020}%
  \BibitemOpen
  \bibfield{author}{%
  \bibinfo {author} {\bibfnamefont{K.}~\bibnamefont{Kandrai}}, \bibinfo
  {author} {\bibfnamefont{G.}~\bibnamefont{Kukucska}}, \bibinfo {author}
  {\bibfnamefont{P.}~\bibnamefont{Vancs{\'o}}}, \bibinfo {author}
  {\bibfnamefont{J.}~\bibnamefont{Koltai}}, \bibinfo {author}
  {\bibfnamefont{G.}~\bibnamefont{Baranka}}, \bibinfo {author}
  {\bibfnamefont{Z.~E.}\ \bibnamefont{Horv{\'a}th}}, \bibinfo {author}
  {\bibfnamefont{{\'A}.}~\bibnamefont{Hoffmann}}, \bibinfo {author}
  {\bibfnamefont{A.}~\bibnamefont{Vymazalov{\'a}}}, \bibinfo {author}
  {\bibfnamefont{L.}~\bibnamefont{Tapaszt{\'o}}},\ and\ \bibinfo {author}
  {\bibfnamefont{P.}~\bibnamefont{Nemes-Incze}},\ }%
  \href{https://pubs.acs.org/doi/10.1021/acs.nanolett.0c01499}{\emph{\bibinfo
  {title} {Signature of large-gap quantum spin Hall state in the layered
  mineral jacutingaite}}},\ \bibinfo {journal} {Nano Letters}\ \textbf{\bibinfo
  {volume} {20}},\ \bibinfo {pages} {5207} (\bibinfo {year} {2020}).~%
  \bibAnnoteFile{Stop}{Kandrai2020}%
\bibitem{Fu2007}%
  \BibitemOpen
  \bibfield{author}{%
  \bibinfo {author} {\bibfnamefont{L.}~\bibnamefont{Fu}}, \bibinfo {author}
  {\bibfnamefont{C.~L.}\ \bibnamefont{Kane}},\ and\ \bibinfo {author}
  {\bibfnamefont{E.~J.}\ \bibnamefont{Mele}},\ }%
  \Doi{10.1103/PhysRevLett.98.106803}{\emph{\bibinfo {title} {Topological
  Insulators in Three Dimensions}}},\ \bibinfo {journal} {Phys. Rev. Lett.}\
  \textbf{\bibinfo {volume} {98}},\ \bibinfo {pages} {106803} (\bibinfo {year}
  {2007}).~%
  \bibAnnoteFile{Stop}{Fu2007}%
\bibitem{Marrazzo2020}%
  \BibitemOpen
  \bibfield{author}{%
  \bibinfo {author} {\bibfnamefont{A.}~\bibnamefont{Marrazzo}}, \bibinfo
  {author} {\bibfnamefont{N.}~\bibnamefont{Marzari}},\ and\ \bibinfo {author}
  {\bibfnamefont{M.}~\bibnamefont{Gibertini}},\ }%
  \Doi{10.1103/PhysRevResearch.2.012063}{\emph{\bibinfo {title} {Emergent dual
  topology in the three-dimensional Kane-Mele
  ${\mathrm{Pt}}_{2}{\mathrm{HgSe}}_{3}$}}},\ \bibinfo {journal} {Phys. Rev.
  Research}\ \textbf{\bibinfo {volume} {2}},\ \bibinfo {pages} {012063}
  (\bibinfo {year} {2020}).~%
  \bibAnnoteFile{Stop}{Marrazzo2020}%
\bibitem{facio_prm_2019}%
  \BibitemOpen
  \bibfield{author}{%
  \bibinfo {author} {\bibfnamefont{J.~I.}\ \bibnamefont{Facio}}, \bibinfo
  {author} {\bibfnamefont{S.~K.}\ \bibnamefont{Das}}, \bibinfo {author}
  {\bibfnamefont{Y.}~\bibnamefont{Zhang}}, \bibinfo {author}
  {\bibfnamefont{K.}~\bibnamefont{Koepernik}}, \bibinfo {author}
  {\bibfnamefont{J.}~\bibnamefont{van~den Brink}},\ and\ \bibinfo {author}
  {\bibfnamefont{I.~C.}\ \bibnamefont{Fulga}},\ }%
  \Doi{10.1103/PhysRevMaterials.3.074202}{\emph{\bibinfo {title} {Dual topology
  in jacutingaite ${\mathrm{Pt}}_{2}{\mathrm{HgSe}}_{3}$}}},\ \bibinfo
  {journal} {Phys. Rev. Materials}\ \textbf{\bibinfo {volume} {3}},\ \bibinfo
  {pages} {074202} (\bibinfo {year} {2019}).~%
  \bibAnnoteFile{Stop}{facio_prm_2019}%
\bibitem{Ghosh2020}%
  \BibitemOpen
  \bibfield{author}{%
  \bibinfo {author} {\bibfnamefont{B.}~\bibnamefont{Ghosh}}, \bibinfo {author}
  {\bibfnamefont{S.}~\bibnamefont{Mardanya}}, \bibinfo {author}
  {\bibfnamefont{B.}~\bibnamefont{Singh}}, \bibinfo {author}
  {\bibfnamefont{X.}~\bibnamefont{Zhou}}, \bibinfo {author}
  {\bibfnamefont{B.}~\bibnamefont{Wang}}, \bibinfo {author}
  {\bibfnamefont{T.-R.}\ \bibnamefont{Chang}}, \bibinfo {author}
  {\bibfnamefont{C.}~\bibnamefont{Su}}, \bibinfo {author}
  {\bibfnamefont{H.}~\bibnamefont{Lin}}, \bibinfo {author}
  {\bibfnamefont{A.}~\bibnamefont{Agarwal}},\ and\ \bibinfo {author}
  {\bibfnamefont{A.}~\bibnamefont{Bansil}},\ }%
  \Doi{10.1103/PhysRevB.100.235101}{\emph{\bibinfo {title} {Saddle-point Van
  Hove singularity and dual topological state in
  ${\mathrm{Pt}}_{2}{\mathrm{HgSe}}_{3}$}}},\ \bibinfo {journal} {Phys. Rev.
  B}\ \textbf{\bibinfo {volume} {100}},\ \bibinfo {pages} {235101} (\bibinfo
  {year} {2019}).~%
  \bibAnnoteFile{Stop}{Ghosh2020}%
\bibitem{Mauro2020}%
  \BibitemOpen
  \bibfield{author}{%
  \bibinfo {author} {\bibfnamefont{D.}~\bibnamefont{Mauro}}, \bibinfo {author}
  {\bibfnamefont{H.}~\bibnamefont{Henck}}, \bibinfo {author}
  {\bibfnamefont{M.}~\bibnamefont{Gibertini}}, \bibinfo {author}
  {\bibfnamefont{M.}~\bibnamefont{Filippone}}, \bibinfo {author}
  {\bibfnamefont{E.}~\bibnamefont{Giannini}}, \bibinfo {author}
  {\bibfnamefont{I.}~\bibnamefont{Guti{\'{e}}rrez-Lezama}},\ and\ \bibinfo
  {author} {\bibfnamefont{A.~F.}\ \bibnamefont{Morpurgo}},\ }%
  \Doi{10.1088/2053-1583/ab7689}{\emph{\bibinfo {title} {Multi-frequency
  Shubnikov-de Haas oscillations in topological semimetal
  $\mathrm{Pt_2HgSe_3}$}}},\ \bibinfo {journal} {2D Materials}\
  \textbf{\bibinfo {volume} {7}},\ \bibinfo {pages} {025042} (\bibinfo {year}
  {2020}).~%
  \bibAnnoteFile{Stop}{Mauro2020}%
\bibitem{Pei2021}%
  \BibitemOpen
  \bibfield{author}{%
  \bibinfo {author} {\bibfnamefont{C.}~\bibnamefont{Pei}}, \bibinfo {author}
  {\bibfnamefont{S.}~\bibnamefont{Jin}}, \bibinfo {author}
  {\bibfnamefont{P.}~\bibnamefont{Huang}}, \bibinfo {author}
  {\bibfnamefont{A.}~\bibnamefont{Vymazalova}}, \bibinfo {author}
  {\bibfnamefont{L.}~\bibnamefont{Gao}}, \bibinfo {author}
  {\bibfnamefont{Y.}~\bibnamefont{Zhao}}, \bibinfo {author}
  {\bibfnamefont{W.}~\bibnamefont{Cao}}, \bibinfo {author}
  {\bibfnamefont{C.}~\bibnamefont{Li}}, \bibinfo {author}
  {\bibfnamefont{P.}~\bibnamefont{Nemes-Incze}}, \bibinfo {author}
  {\bibfnamefont{Y.}~\bibnamefont{Chen}}, \bibinfo {author}
  {\bibfnamefont{H.}~\bibnamefont{Liu}}, \bibinfo {author}
  {\bibfnamefont{G.}~\bibnamefont{Li}},\ and\ \bibinfo {author}
  {\bibfnamefont{Y.}~\bibnamefont{Qi}},\ }%
  \href{https://arxiv.org/abs/2102.08801}{\emph{\bibinfo {title}
  {Pressure-induced Superconductivity in dual-topological semimetal
  $\mathrm{Pt_2HgSe_3}$}}},\ \bibinfo {journal} {arXiv:2102.08801}\  (\bibinfo
  {year} {2021}).~%
  \bibAnnoteFile{Stop}{Pei2021}%
\bibitem{Cucchi2020}%
  \BibitemOpen
  \bibfield{author}{%
  \bibinfo {author} {\bibfnamefont{I.}~\bibnamefont{Cucchi}}, \bibinfo {author}
  {\bibfnamefont{A.}~\bibnamefont{Marrazzo}}, \bibinfo {author}
  {\bibfnamefont{E.}~\bibnamefont{Cappelli}}, \bibinfo {author}
  {\bibfnamefont{S.}~\bibnamefont{Ricc\`o}}, \bibinfo {author}
  {\bibfnamefont{F.~Y.}\ \bibnamefont{Bruno}}, \bibinfo {author}
  {\bibfnamefont{S.}~\bibnamefont{Lisi}}, \bibinfo {author}
  {\bibfnamefont{M.}~\bibnamefont{Hoesch}}, \bibinfo {author}
  {\bibfnamefont{T.~K.}\ \bibnamefont{Kim}}, \bibinfo {author}
  {\bibfnamefont{C.}~\bibnamefont{Cacho}}, \bibinfo {author}
  {\bibfnamefont{C.}~\bibnamefont{Besnard}}, \bibinfo {author}
  {\bibfnamefont{E.}~\bibnamefont{Giannini}}, \bibinfo {author}
  {\bibfnamefont{N.}~\bibnamefont{Marzari}}, \bibinfo {author}
  {\bibfnamefont{M.}~\bibnamefont{Gibertini}}, \bibinfo {author}
  {\bibfnamefont{F.}~\bibnamefont{Baumberger}},\ and\ \bibinfo {author}
  {\bibfnamefont{A.}~\bibnamefont{Tamai}},\ }%
  \Doi{10.1103/PhysRevLett.124.106402}{\emph{\bibinfo {title} {Bulk and Surface
  Electronic Structure of the Dual-Topology Semimetal
  ${\mathrm{Pt}}_{2}{\mathrm{HgSe}}_{3}$}}},\ \bibinfo {journal} {Phys. Rev.
  Lett.}\ \textbf{\bibinfo {volume} {124}},\ \bibinfo {pages} {106402}
  (\bibinfo {year} {2020}).~%
  \bibAnnoteFile{Stop}{Cucchi2020}%
\bibitem{QE-2009}%
  \BibitemOpen
  \bibfield{author}{%
  \bibinfo {author} {\bibfnamefont{P.}~\bibnamefont{Giannozzi}}, \bibinfo
  {author} {\bibfnamefont{S.}~\bibnamefont{Baroni}}, \bibinfo {author}
  {\bibfnamefont{N.}~\bibnamefont{Bonini}}, \bibinfo {author}
  {\bibfnamefont{M.}~\bibnamefont{Calandra}}, \bibinfo {author}
  {\bibfnamefont{R.}~\bibnamefont{Car}}, \bibinfo {author}
  {\bibfnamefont{C.}~\bibnamefont{Cavazzoni}}, \bibinfo {author}
  {\bibfnamefont{D.}~\bibnamefont{Ceresoli}}, \bibinfo {author}
  {\bibfnamefont{G.~L.}\ \bibnamefont{Chiarotti}}, \bibinfo {author}
  {\bibfnamefont{M.}~\bibnamefont{Cococcioni}}, \bibinfo {author}
  {\bibfnamefont{I.}~\bibnamefont{Dabo}}, \bibinfo {author}
  {\bibfnamefont{A.}~\bibnamefont{{Dal Corso}}}, \bibinfo {author}
  {\bibfnamefont{S.}~\bibnamefont{de~Gironcoli}}, \bibinfo {author}
  {\bibfnamefont{S.}~\bibnamefont{Fabris}}, \bibinfo {author}
  {\bibfnamefont{G.}~\bibnamefont{Fratesi}}, \bibinfo {author}
  {\bibfnamefont{R.}~\bibnamefont{Gebauer}}, \bibinfo {author}
  {\bibfnamefont{U.}~\bibnamefont{Gerstmann}}, \bibinfo {author}
  {\bibfnamefont{C.}~\bibnamefont{Gougoussis}}, \bibinfo {author}
  {\bibfnamefont{A.}~\bibnamefont{Kokalj}}, \bibinfo {author}
  {\bibfnamefont{M.}~\bibnamefont{Lazzeri}}, \bibinfo {author}
  {\bibfnamefont{L.}~\bibnamefont{Martin-Samos}}, \bibinfo {author}
  {\bibfnamefont{N.}~\bibnamefont{Marzari}}, \bibinfo {author}
  {\bibfnamefont{F.}~\bibnamefont{Mauri}}, \bibinfo {author}
  {\bibfnamefont{R.}~\bibnamefont{Mazzarello}}, \bibinfo {author}
  {\bibfnamefont{S.}~\bibnamefont{Paolini}}, \bibinfo {author}
  {\bibfnamefont{A.}~\bibnamefont{Pasquarello}}, \bibinfo {author}
  {\bibfnamefont{L.}~\bibnamefont{Paulatto}}, \bibinfo {author}
  {\bibfnamefont{C.}~\bibnamefont{Sbraccia}}, \bibinfo {author}
  {\bibfnamefont{S.}~\bibnamefont{Scandolo}}, \bibinfo {author}
  {\bibfnamefont{G.}~\bibnamefont{Sclauzero}}, \bibinfo {author}
  {\bibfnamefont{A.~P.}\ \bibnamefont{Seitsonen}}, \bibinfo {author}
  {\bibfnamefont{A.}~\bibnamefont{Smogunov}}, \bibinfo {author}
  {\bibfnamefont{P.}~\bibnamefont{Umari}},\ and\ \bibinfo {author}
  {\bibfnamefont{R.~M.}\ \bibnamefont{Wentzcovitch}},\ }%
  \href{http://www.quantum-espresso.org}{\emph{\bibinfo {title} {QUANTUM
  ESPRESSO: a modular and open-source software project for quantum simulations
  of materials}}},\ \bibinfo {journal} {Journal of Physics: Condensed Matter}\
  \textbf{\bibinfo {volume} {21}},\ \bibinfo {pages} {395502 (19pp)} (\bibinfo
  {year} {2009}).~%
  \bibAnnoteFile{Stop}{QE-2009}%
\bibitem{QE-2017}%
  \BibitemOpen
  \bibfield{author}{%
  \bibinfo {author} {\bibfnamefont{P.}~\bibnamefont{Giannozzi}}, \bibinfo
  {author} {\bibfnamefont{O.}~\bibnamefont{Andreussi}}, \bibinfo {author}
  {\bibfnamefont{T.}~\bibnamefont{Brumme}}, \bibinfo {author}
  {\bibfnamefont{O.}~\bibnamefont{Bunau}}, \bibinfo {author}
  {\bibfnamefont{M.~B.}\ \bibnamefont{Nardelli}}, \bibinfo {author}
  {\bibfnamefont{M.}~\bibnamefont{Calandra}}, \bibinfo {author}
  {\bibfnamefont{R.}~\bibnamefont{Car}}, \bibinfo {author}
  {\bibfnamefont{C.}~\bibnamefont{Cavazzoni}}, \bibinfo {author}
  {\bibfnamefont{D.}~\bibnamefont{Ceresoli}}, \bibinfo {author}
  {\bibfnamefont{M.}~\bibnamefont{Cococcioni}}, \bibinfo {author}
  {\bibfnamefont{N.}~\bibnamefont{Colonna}}, \bibinfo {author}
  {\bibfnamefont{I.}~\bibnamefont{Carnimeo}}, \bibinfo {author}
  {\bibfnamefont{A.~D.}\ \bibnamefont{Corso}}, \bibinfo {author}
  {\bibfnamefont{S.}~\bibnamefont{de~Gironcoli}}, \bibinfo {author}
  {\bibfnamefont{P.}~\bibnamefont{Delugas}}, \bibinfo {author}
  {\bibfnamefont{R.~A.~D.}\ \bibnamefont{Jr}}, \bibinfo {author}
  {\bibfnamefont{A.}~\bibnamefont{Ferretti}}, \bibinfo {author}
  {\bibfnamefont{A.}~\bibnamefont{Floris}}, \bibinfo {author}
  {\bibfnamefont{G.}~\bibnamefont{Fratesi}}, \bibinfo {author}
  {\bibfnamefont{G.}~\bibnamefont{Fugallo}}, \bibinfo {author}
  {\bibfnamefont{R.}~\bibnamefont{Gebauer}}, \bibinfo {author}
  {\bibfnamefont{U.}~\bibnamefont{Gerstmann}}, \bibinfo {author}
  {\bibfnamefont{F.}~\bibnamefont{Giustino}}, \bibinfo {author}
  {\bibfnamefont{T.}~\bibnamefont{Gorni}}, \bibinfo {author}
  {\bibfnamefont{J.}~\bibnamefont{Jia}}, \bibinfo {author}
  {\bibfnamefont{M.}~\bibnamefont{Kawamura}}, \bibinfo {author}
  {\bibfnamefont{H.-Y.}\ \bibnamefont{Ko}}, \bibinfo {author}
  {\bibfnamefont{A.}~\bibnamefont{Kokalj}}, \bibinfo {author}
  {\bibfnamefont{E.}~\bibnamefont{K{\"u}{\c c}{\"u}kbenli}}, \bibinfo {author}
  {\bibfnamefont{M.}~\bibnamefont{Lazzeri}}, \bibinfo {author}
  {\bibfnamefont{M.}~\bibnamefont{Marsili}}, \bibinfo {author}
  {\bibfnamefont{N.}~\bibnamefont{Marzari}}, \bibinfo {author}
  {\bibfnamefont{F.}~\bibnamefont{Mauri}}, \bibinfo {author}
  {\bibfnamefont{N.~L.}\ \bibnamefont{Nguyen}}, \bibinfo {author}
  {\bibfnamefont{H.-V.}\ \bibnamefont{Nguyen}}, \bibinfo {author}
  {\bibfnamefont{A.~O.}\ \bibnamefont{de-la Roza}}, \bibinfo {author}
  {\bibfnamefont{L.}~\bibnamefont{Paulatto}}, \bibinfo {author}
  {\bibfnamefont{S.}~\bibnamefont{Ponc{\'e}}}, \bibinfo {author}
  {\bibfnamefont{D.}~\bibnamefont{Rocca}}, \bibinfo {author}
  {\bibfnamefont{R.}~\bibnamefont{Sabatini}}, \bibinfo {author}
  {\bibfnamefont{B.}~\bibnamefont{Santra}}, \bibinfo {author}
  {\bibfnamefont{M.}~\bibnamefont{Schlipf}}, \bibinfo {author}
  {\bibfnamefont{A.~P.}\ \bibnamefont{Seitsonen}}, \bibinfo {author}
  {\bibfnamefont{A.}~\bibnamefont{Smogunov}}, \bibinfo {author}
  {\bibfnamefont{I.}~\bibnamefont{Timrov}}, \bibinfo {author}
  {\bibfnamefont{T.}~\bibnamefont{Thonhauser}}, \bibinfo {author}
  {\bibfnamefont{P.}~\bibnamefont{Umari}}, \bibinfo {author}
  {\bibfnamefont{N.}~\bibnamefont{Vast}}, \bibinfo {author}
  {\bibfnamefont{X.}~\bibnamefont{Wu}},\ and\ \bibinfo {author}
  {\bibfnamefont{S.}~\bibnamefont{Baroni}},\ }%
  \href{http://stacks.iop.org/0953-8984/29/i=46/a=465901}{\emph{\bibinfo
  {title} {Advanced capabilities for materials modelling with QUANTUM
  ESPRESSO}}},\ \bibinfo {journal} {Journal of Physics: Condensed Matter}\
  \textbf{\bibinfo {volume} {29}},\ \bibinfo {pages} {465901} (\bibinfo {year}
  {2017}).~%
  \bibAnnoteFile{Stop}{QE-2017}%
\bibitem{Sohier2017}%
  \BibitemOpen
  \bibfield{author}{%
  \bibinfo {author} {\bibfnamefont{T.}~\bibnamefont{Sohier}}, \bibinfo {author}
  {\bibfnamefont{M.}~\bibnamefont{Calandra}},\ and\ \bibinfo {author}
  {\bibfnamefont{F.}~\bibnamefont{Mauri}},\ }%
  \Doi{10.1103/PhysRevB.96.075448}{\emph{\bibinfo {title} {Density functional
  perturbation theory for gated two-dimensional heterostructures: Theoretical
  developments and application to flexural phonons in graphene}}},\ \bibinfo
  {journal} {Phys. Rev. B}\ \textbf{\bibinfo {volume} {96}},\ \bibinfo {pages}
  {075448} (\bibinfo {year} {2017}).~%
  \bibAnnoteFile{Stop}{Sohier2017}%
\bibitem{Dion2004}%
  \BibitemOpen
  \bibfield{author}{%
  \bibinfo {author} {\bibfnamefont{M.}~\bibnamefont{Dion}}, \bibinfo {author}
  {\bibfnamefont{H.}~\bibnamefont{Rydberg}}, \bibinfo {author}
  {\bibfnamefont{E.}~\bibnamefont{Schr\"oder}}, \bibinfo {author}
  {\bibfnamefont{D.~C.}\ \bibnamefont{Langreth}},\ and\ \bibinfo {author}
  {\bibfnamefont{B.~I.}\ \bibnamefont{Lundqvist}},\ }%
  \Doi{10.1103/PhysRevLett.92.246401}{\emph{\bibinfo {title} {Van der Waals
  Density Functional for General Geometries}}},\ \bibinfo {journal} {Phys. Rev.
  Lett.}\ \textbf{\bibinfo {volume} {92}},\ \bibinfo {pages} {246401} (\bibinfo
  {year} {2004}).~%
  \bibAnnoteFile{Stop}{Dion2004}%
\bibitem{Berland2014}%
  \BibitemOpen
  \bibfield{author}{%
  \bibinfo {author} {\bibfnamefont{K.}~\bibnamefont{Berland}}\ and\ \bibinfo
  {author} {\bibfnamefont{P.}~\bibnamefont{Hyldgaard}},\ }%
  \Doi{10.1103/PhysRevB.89.035412}{\emph{\bibinfo {title} {Exchange functional
  that tests the robustness of the plasmon description of the van der Waals
  density functional}}},\ \bibinfo {journal} {Phys. Rev. B}\ \textbf{\bibinfo
  {volume} {89}},\ \bibinfo {pages} {035412} (\bibinfo {year} {2014}).~%
  \bibAnnoteFile{Stop}{Berland2014}%
\bibitem{Berland2015}%
  \BibitemOpen
  \bibfield{author}{%
  \bibinfo {author} {\bibfnamefont{K.}~\bibnamefont{Berland}}, \bibinfo
  {author} {\bibfnamefont{V.~R.}\ \bibnamefont{Cooper}}, \bibinfo {author}
  {\bibfnamefont{K.}~\bibnamefont{Lee}}, \bibinfo {author}
  {\bibfnamefont{E.}~\bibnamefont{Schr{\"o}der}}, \bibinfo {author}
  {\bibfnamefont{T.}~\bibnamefont{Thonhauser}}, \bibinfo {author}
  {\bibfnamefont{P.}~\bibnamefont{Hyldgaard}},\ and\ \bibinfo {author}
  {\bibfnamefont{B.~I.}\ \bibnamefont{Lundqvist}},\ }%
  \Doi{10.1088/0034-4885/78/6/066501}{\emph{\bibinfo {title} {{van der Waals
  forces in density functional theory: a review of the vdW-DF method}}}},\
  \bibinfo {journal} {Rep. Prog. Phys.}\ \textbf{\bibinfo {volume} {78}},\
  \bibinfo {pages} {066501} (\bibinfo {year} {2015}).~%
  \bibAnnoteFile{Stop}{Berland2015}%
\bibitem{Borlido2019}%
  \BibitemOpen
  \bibfield{author}{%
  \bibinfo {author} {\bibfnamefont{P.}~\bibnamefont{Borlido}}, \bibinfo
  {author} {\bibfnamefont{T.}~\bibnamefont{Aull}}, \bibinfo {author}
  {\bibfnamefont{A.~W.}\ \bibnamefont{Huran}}, \bibinfo {author}
  {\bibfnamefont{F.}~\bibnamefont{Tran}}, \bibinfo {author}
  {\bibfnamefont{M.~A.~L.}\ \bibnamefont{Marques}},\ and\ \bibinfo {author}
  {\bibfnamefont{S.}~\bibnamefont{Botti}},\ }%
  \Doi{10.1021/acs.jctc.9b00322}{\emph{\bibinfo {title} {Large-Scale Benchmark
  of Exchange--Correlation Functionals for the Determination of Electronic Band
  Gaps of Solids}}},\ \bibinfo {journal} {Journal of Chemical Theory and
  Computation}\ \textbf{\bibinfo {volume} {15}},\ \bibinfo {pages} {5069}
  (\bibinfo {year} {2019}).~%
  \bibAnnoteFile{Stop}{Borlido2019}%
\bibitem{wannier_review_12}%
  \BibitemOpen
  \bibfield{author}{%
  \bibinfo {author} {\bibfnamefont{N.}~\bibnamefont{Marzari}}, \bibinfo
  {author} {\bibfnamefont{A.~A.}\ \bibnamefont{Mostofi}}, \bibinfo {author}
  {\bibfnamefont{J.~R.}\ \bibnamefont{Yates}}, \bibinfo {author}
  {\bibfnamefont{I.}~\bibnamefont{Souza}},\ and\ \bibinfo {author}
  {\bibfnamefont{D.}~\bibnamefont{Vanderbilt}},\ }%
  \Doi{10.1103/RevModPhys.84.1419}{\emph{\bibinfo {title} {Maximally Localized
  {{Wannier}} Functions: {{Theory}} and Applications}}},\ \bibinfo {journal}
  {Reviews of Modern Physics}\ \textbf{\bibinfo {volume} {84}},\ \bibinfo
  {pages} {1419} (\bibinfo {year} {2012}).~%
  \bibAnnoteFile{Stop}{wannier_review_12}%
\bibitem{Pizzi2020}%
  \BibitemOpen
  \bibfield{author}{%
  \bibinfo {author} {\bibfnamefont{G.}~\bibnamefont{Pizzi}}, \bibinfo {author}
  {\bibfnamefont{V.}~\bibnamefont{Vitale}}, \bibinfo {author}
  {\bibfnamefont{R.}~\bibnamefont{Arita}}, \bibinfo {author}
  {\bibfnamefont{S.}~\bibnamefont{Bl{\"u}gel}}, \bibinfo {author}
  {\bibfnamefont{F.}~\bibnamefont{Freimuth}}, \bibinfo {author}
  {\bibfnamefont{G.}~\bibnamefont{G{\'{e}}ranton}}, \bibinfo {author}
  {\bibfnamefont{M.}~\bibnamefont{Gibertini}}, \bibinfo {author}
  {\bibfnamefont{D.}~\bibnamefont{Gresch}}, \bibinfo {author}
  {\bibfnamefont{C.}~\bibnamefont{Johnson}}, \bibinfo {author}
  {\bibfnamefont{T.}~\bibnamefont{Koretsune}}, \bibinfo {author}
  {\bibfnamefont{J.}~\bibnamefont{Iba{\~{n}}ez-Azpiroz}}, \bibinfo {author}
  {\bibfnamefont{H.}~\bibnamefont{Lee}}, \bibinfo {author}
  {\bibfnamefont{J.-M.}\ \bibnamefont{Lihm}}, \bibinfo {author}
  {\bibfnamefont{D.}~\bibnamefont{Marchand}}, \bibinfo {author}
  {\bibfnamefont{A.}~\bibnamefont{Marrazzo}}, \bibinfo {author}
  {\bibfnamefont{Y.}~\bibnamefont{Mokrousov}}, \bibinfo {author}
  {\bibfnamefont{J.~I.}\ \bibnamefont{Mustafa}}, \bibinfo {author}
  {\bibfnamefont{Y.}~\bibnamefont{Nohara}}, \bibinfo {author}
  {\bibfnamefont{Y.}~\bibnamefont{Nomura}}, \bibinfo {author}
  {\bibfnamefont{L.}~\bibnamefont{Paulatto}}, \bibinfo {author}
  {\bibfnamefont{S.}~\bibnamefont{Ponc{\'{e}}}}, \bibinfo {author}
  {\bibfnamefont{T.}~\bibnamefont{Ponweiser}}, \bibinfo {author}
  {\bibfnamefont{J.}~\bibnamefont{Qiao}}, \bibinfo {author}
  {\bibfnamefont{F.}~\bibnamefont{Th{\"o}le}}, \bibinfo {author}
  {\bibfnamefont{S.~S.}\ \bibnamefont{Tsirkin}}, \bibinfo {author}
  {\bibfnamefont{M.}~\bibnamefont{Wierzbowska}}, \bibinfo {author}
  {\bibfnamefont{N.}~\bibnamefont{Marzari}}, \bibinfo {author}
  {\bibfnamefont{D.}~\bibnamefont{Vanderbilt}}, \bibinfo {author}
  {\bibfnamefont{I.}~\bibnamefont{Souza}}, \bibinfo {author}
  {\bibfnamefont{A.~A.}\ \bibnamefont{Mostofi}},\ and\ \bibinfo {author}
  {\bibfnamefont{J.~R.}\ \bibnamefont{Yates}},\ }%
  \Doi{10.1088/1361-648x/ab51ff}{\emph{\bibinfo {title} {Wannier90 as a
  community code: new features and applications}}},\ \bibinfo {journal}
  {Journal of Physics: Condensed Matter}\ \textbf{\bibinfo {volume} {32}},\
  \bibinfo {pages} {165902} (\bibinfo {year} {2020}).~%
  \bibAnnoteFile{Stop}{Pizzi2020}%
\bibitem{wannier_tools_18}%
  \BibitemOpen
  \bibfield{author}{%
  \bibinfo {author} {\bibfnamefont{Q.}~\bibnamefont{Wu}}, \bibinfo {author}
  {\bibfnamefont{S.}~\bibnamefont{Zhang}}, \bibinfo {author}
  {\bibfnamefont{H.-F.}\ \bibnamefont{Song}}, \bibinfo {author}
  {\bibfnamefont{M.}~\bibnamefont{Troyer}},\ and\ \bibinfo {author}
  {\bibfnamefont{A.~A.}\ \bibnamefont{Soluyanov}},\ }%
  \Doi{https://doi.org/10.1016/j.cpc.2017.09.033}{\emph{\bibinfo {title}
  {{WannierTools}: {An} open-source software package for novel topological
  materials}}},\ \bibinfo {journal} {Computer Physics Communications}\
  \textbf{\bibinfo {volume} {224}},\ \bibinfo {pages} {405 } (\bibinfo {year}
  {2018}).~%
  \bibAnnoteFile{Stop}{wannier_tools_18}%
\bibitem{Soluyanov2011}%
  \BibitemOpen
  \bibfield{author}{%
  \bibinfo {author} {\bibfnamefont{A.~A.}\ \bibnamefont{Soluyanov}}\ and\
  \bibinfo {author} {\bibfnamefont{D.}~\bibnamefont{Vanderbilt}},\ }%
  \Doi{10.1103/PhysRevB.83.235401}{\emph{\bibinfo {title} {Computing
  Topological Invariants without Inversion Symmetry}}},\ \bibinfo {journal}
  {Physical Review B}\ \textbf{\bibinfo {volume} {83}},\ \bibinfo {pages}
  {235401} (\bibinfo {year} {2011}).~%
  \bibAnnoteFile{Stop}{Soluyanov2011}%
\bibitem{Yu2011}%
  \BibitemOpen
  \bibfield{author}{%
  \bibinfo {author} {\bibfnamefont{R.}~\bibnamefont{Yu}}, \bibinfo {author}
  {\bibfnamefont{X.~L.}\ \bibnamefont{Qi}}, \bibinfo {author}
  {\bibfnamefont{A.}~\bibnamefont{Bernevig}}, \bibinfo {author}
  {\bibfnamefont{Z.}~\bibnamefont{Fang}},\ and\ \bibinfo {author}
  {\bibfnamefont{X.}~\bibnamefont{Dai}},\ }%
  \Doi{10.1103/PhysRevB.84.075119}{\emph{\bibinfo {title} {Equivalent
  expression of ${\mathbb{Z}}_{2}$ topological invariant for band insulators
  using the non-Abelian Berry connection}}},\ \bibinfo {journal} {Phys. Rev.
  B}\ \textbf{\bibinfo {volume} {84}},\ \bibinfo {pages} {075119} (\bibinfo
  {year} {2011}).~%
  \bibAnnoteFile{Stop}{Yu2011}%
\bibitem{Sgiarovello2001}%
  \BibitemOpen
  \bibfield{author}{%
  \bibinfo {author} {\bibfnamefont{C.}~\bibnamefont{Sgiarovello}}, \bibinfo
  {author} {\bibfnamefont{M.}~\bibnamefont{Peressi}},\ and\ \bibinfo {author}
  {\bibfnamefont{R.}~\bibnamefont{Resta}},\ }%
  \Doi{10.1103/PhysRevB.64.115202}{\emph{\bibinfo {title} {Electron
  localization in the insulating state: Application to crystalline
  semiconductors}}},\ \bibinfo {journal} {Phys. Rev. B}\ \textbf{\bibinfo
  {volume} {64}},\ \bibinfo {pages} {115202} (\bibinfo {year} {2001}).~%
  \bibAnnoteFile{Stop}{Sgiarovello2001}%
\bibitem{fu_topological_2007}%
  \BibitemOpen
  \bibfield{author}{%
  \bibinfo {author} {\bibfnamefont{L.}~\bibnamefont{Fu}}\ and\ \bibinfo
  {author} {\bibfnamefont{C.~L.}\ \bibnamefont{Kane}},\ }%
  \Doi{10.1103/PhysRevB.76.045302}{\emph{\bibinfo {title} {Topological
  Insulators with Inversion Symmetry}}},\ \bibinfo {journal} {Physical Review
  B}\ \textbf{\bibinfo {volume} {76}},\ \bibinfo {pages} {045302} (\bibinfo
  {year} {2007}).~%
  \bibAnnoteFile{Stop}{fu_topological_2007}%
\bibitem{liu_buckledSOC_11}%
  \BibitemOpen
  \bibfield{author}{%
  \bibinfo {author} {\bibfnamefont{C.-C.}\ \bibnamefont{Liu}}, \bibinfo
  {author} {\bibfnamefont{H.}~\bibnamefont{Jiang}},\ and\ \bibinfo {author}
  {\bibfnamefont{Y.}~\bibnamefont{Yao}},\ }%
  \Doi{10.1103/PhysRevB.84.195430}{\emph{\bibinfo {title} {Low-Energy Effective
  {{Hamiltonian}} Involving Spin-Orbit Coupling in Silicene and Two-Dimensional
  Germanium and Tin}}},\ \bibinfo {journal} {Physical Review B}\
  \textbf{\bibinfo {volume} {84}},\ \bibinfo {pages} {195430} (\bibinfo {year}
  {2011}).~%
  \bibAnnoteFile{Stop}{liu_buckledSOC_11}%
\bibitem{Wu2019}%
  \BibitemOpen
  \bibfield{author}{%
  \bibinfo {author} {\bibfnamefont{X.}~\bibnamefont{Wu}}, \bibinfo {author}
  {\bibfnamefont{M.}~\bibnamefont{Fink}}, \bibinfo {author}
  {\bibfnamefont{W.}~\bibnamefont{Hanke}}, \bibinfo {author}
  {\bibfnamefont{R.}~\bibnamefont{Thomale}},\ and\ \bibinfo {author}
  {\bibfnamefont{D.}~\bibnamefont{Di~Sante}},\ }%
  \Doi{10.1103/PhysRevB.100.041117}{\emph{\bibinfo {title} {Unconventional
  superconductivity in a doped quantum spin Hall insulator}}},\ \bibinfo
  {journal} {Phys. Rev. B}\ \textbf{\bibinfo {volume} {100}},\ \bibinfo {pages}
  {041117(R)} (\bibinfo {year} {2019}).~%
  \bibAnnoteFile{Stop}{Wu2019}%
\bibitem{perdew_pbe_96}%
  \BibitemOpen
  \bibfield{author}{%
  \bibinfo {author} {\bibfnamefont{J.~P.}\ \bibnamefont{Perdew}}, \bibinfo
  {author} {\bibfnamefont{K.}~\bibnamefont{Burke}},\ and\ \bibinfo {author}
  {\bibfnamefont{M.}~\bibnamefont{Ernzerhof}},\ }%
  \Doi{10.1103/PhysRevLett.77.3865}{\emph{\bibinfo {title} {Generalized
  {{Gradient Approximation Made Simple}}}}},\ \bibinfo {journal} {Physical
  Review Letters}\ \textbf{\bibinfo {volume} {77}},\ \bibinfo {pages} {3865}
  (\bibinfo {year} {1996}).~%
  \bibAnnoteFile{Stop}{perdew_pbe_96}%
\bibitem{HSE}%
  \BibitemOpen
  \bibfield{author}{%
  \bibinfo {author} {\bibfnamefont{J.}~\bibnamefont{Heyd}}, \bibinfo {author}
  {\bibfnamefont{G.~E.}\ \bibnamefont{Scuseria}},\ and\ \bibinfo {author}
  {\bibfnamefont{M.}~\bibnamefont{Ernzerhof}},\ }%
  \Doi{10.1063/1.1564060}{\emph{\bibinfo {title} {Hybrid functionals based on a
  screened Coulomb potential}}},\ \bibinfo {journal} {The Journal of Chemical
  Physics}\ \textbf{\bibinfo {volume} {118}},\ \bibinfo {pages} {8207}
  (\bibinfo {year} {2003}).~%
  \bibAnnoteFile{Stop}{HSE}%
\bibitem{prandini_precision_2018}%
  \BibitemOpen
  \bibfield{author}{%
  \bibinfo {author} {\bibfnamefont{G.}~\bibnamefont{Prandini}}, \bibinfo
  {author} {\bibfnamefont{A.}~\bibnamefont{Marrazzo}}, \bibinfo {author}
  {\bibfnamefont{I.~E.}\ \bibnamefont{Castelli}}, \bibinfo {author}
  {\bibfnamefont{N.}~\bibnamefont{Mounet}},\ and\ \bibinfo {author}
  {\bibfnamefont{N.}~\bibnamefont{Marzari}},\ }%
  \Doi{10.1038/s41524-018-0127-2}{\emph{\bibinfo {title} {Precision and
  efficiency in solid-state pseudopotential calculations}}},\ \bibinfo
  {journal} {npj Computational Materials}\ \textbf{\bibinfo {volume} {4}},\
  \bibinfo {pages} {72} (\bibinfo {year} {2018}).~%
  \bibAnnoteFile{Stop}{prandini_precision_2018}%
\bibitem{mv_smearing_99}%
  \BibitemOpen
  \bibfield{author}{%
  \bibinfo {author} {\bibfnamefont{N.}~\bibnamefont{Marzari}}, \bibinfo
  {author} {\bibfnamefont{D.}~\bibnamefont{Vanderbilt}}, \bibinfo {author}
  {\bibfnamefont{A.}~\bibnamefont{De~Vita}},\ and\ \bibinfo {author}
  {\bibfnamefont{M.~C.}\ \bibnamefont{Payne}},\ }%
  \Doi{10.1103/PhysRevLett.82.3296}{\emph{\bibinfo {title} {Thermal Contraction
  and Disordering of the Al(110) Surface}}},\ \bibinfo {journal} {Phys. Rev.
  Lett.}\ \textbf{\bibinfo {volume} {82}},\ \bibinfo {pages} {3296} (\bibinfo
  {year} {1999}).~%
  \bibAnnoteFile{Stop}{mv_smearing_99}%
\bibitem{dojo_paper_18}%
  \BibitemOpen
  \bibfield{author}{%
  \bibinfo {author} {\bibfnamefont{M.~J.}\ \bibnamefont{van Setten}}, \bibinfo
  {author} {\bibfnamefont{M.}~\bibnamefont{Giantomassi}}, \bibinfo {author}
  {\bibfnamefont{E.}~\bibnamefont{Bousquet}}, \bibinfo {author}
  {\bibfnamefont{M.~J.}\ \bibnamefont{Verstraete}}, \bibinfo {author}
  {\bibfnamefont{D.~R.}\ \bibnamefont{Hamann}}, \bibinfo {author}
  {\bibfnamefont{X.}~\bibnamefont{Gonze}},\ and\ \bibinfo {author}
  {\bibfnamefont{G.~M.}\ \bibnamefont{Rignanese}},\ }%
  \Doi{10.1016/j.cpc.2018.01.012}{\emph{\bibinfo {title} {The {PseudoDojo}:
  {Training} and grading a 85 element optimized norm-conserving pseudopotential
  table}}},\ \bibinfo {journal} {Computer Physics Communications}\
  \textbf{\bibinfo {volume} {226}},\ \bibinfo {pages} {39} (\bibinfo {year}
  {2018}).~%
  \bibAnnoteFile{Stop}{dojo_paper_18}%
\bibitem{hamann_oncv_13}%
  \BibitemOpen
  \bibfield{author}{%
  \bibinfo {author} {\bibfnamefont{D.~R.}\ \bibnamefont{Hamann}},\ }%
  \Doi{10.1103/PhysRevB.88.085117}{\emph{\bibinfo {title} {Optimized
  Norm-Conserving {{Vanderbilt}} Pseudopotentials}}},\ \bibinfo {journal}
  {Physical Review B}\ \textbf{\bibinfo {volume} {88}},\ \bibinfo {pages}
  {085117} (\bibinfo {year} {2013}).~%
  \bibAnnoteFile{Stop}{hamann_oncv_13}%
\bibitem{Schlipf2015}%
  \BibitemOpen
  \bibfield{author}{%
  \bibinfo {author} {\bibfnamefont{M.}~\bibnamefont{Schlipf}}\ and\ \bibinfo
  {author} {\bibfnamefont{F.}~\bibnamefont{Gygi}},\ }%
  \Doi{https://doi.org/10.1016/j.cpc.2015.05.011}{\emph{\bibinfo {title}
  {Optimization algorithm for the generation of ONCV pseudopotentials}}},\
  \bibinfo {journal} {Computer Physics Communications}\ \textbf{\bibinfo
  {volume} {196}},\ \bibinfo {pages} {36 } (\bibinfo {year} {2015}).~%
  \bibAnnoteFile{Stop}{Schlipf2015}%
\bibitem{Scherpelz2016}%
  \BibitemOpen
  \bibfield{author}{%
  \bibinfo {author} {\bibfnamefont{P.}~\bibnamefont{Scherpelz}}, \bibinfo
  {author} {\bibfnamefont{M.}~\bibnamefont{Govoni}}, \bibinfo {author}
  {\bibfnamefont{I.}~\bibnamefont{Hamada}},\ and\ \bibinfo {author}
  {\bibfnamefont{G.}~\bibnamefont{Galli}},\ }%
  \Doi{10.1021/acs.jctc.6b00114}{\emph{\bibinfo {title} {Implementation and
  Validation of Fully Relativistic GW Calculations: Spin--Orbit Coupling in
  Molecules, Nanocrystals, and Solids}}},\ \bibinfo {journal} {J. Chem. Theory
  Comput.}\ \textbf{\bibinfo {volume} {12}},\ \bibinfo {pages} {3523} (\bibinfo
  {year} {2016}).~%
  \bibAnnoteFile{Stop}{Scherpelz2016}%
\bibitem{VESTA}%
  \BibitemOpen
  \bibfield{author}{%
  \bibinfo {author} {\bibfnamefont{K.}~\bibnamefont{Momma}}\ and\ \bibinfo
  {author} {\bibfnamefont{F.}~\bibnamefont{Izumi}},\ }%
  \Doi{10.1107/S0021889811038970}{\emph{\bibinfo {title} {{{\it VESTA3} for
  three-dimensional visualization of crystal, volumetric and morphology
  data}}}},\ \bibinfo {journal} {Journal of Applied Crystallography}\
  \textbf{\bibinfo {volume} {44}},\ \bibinfo {pages} {1272} (\bibinfo {year}
  {2011}).~%
  \bibAnnoteFile{Stop}{VESTA}%
\end{thebibliography}
\end{document}